\documentclass{jfm}
\usepackage[utf8]{inputenc}
\usepackage{amsmath}
\usepackage{amssymb}
\usepackage{physics}
\usepackage{appendix}
\usepackage{graphicx}
\graphicspath{ {././Images/} , {./figure/}, {./Figure}}

\usepackage{float}
\usepackage[caption = true]{subfig}
\usepackage{hyperref}
\hypersetup{
  colorlinks=true,
  linkcolor=blue,
  citecolor=blue,
  filecolor=magenta,
  urlcolor=cyan
}

\usepackage{xcolor}
\def\drawline#1#2{\raise 2.5pt\vbox{\hrule width #1pt height #2pt}}
\def\spacce#1{\hskip #1pt}
\def\solid{\drawline{24}{.5}\nobreak\ }
\def\bdash{\hbox{\drawline{4}{.5}\spacce{2}}}
\def\dashed{\bdash\bdash\bdash\bdash\nobreak\ }
\def\bdot{\hbox{\drawline{1}{.5}\spacce{2}}}
\def\dotted{\hbox{\leaders\bdot\hskip 24pt}\nobreak\ }
\def\chndash{\hbox {\drawline{8.5}{.5}\spacce{2}\drawline{3}{.5}\spacce{2}\drawline{8.5}{.5}}\nobreak\ }

\DeclareMathOperator*{\argmax}{arg\,max} 
\newcommand{\rulesep}{\unskip\ \vrule\ }

\shorttitle{Convection between spherical shells}
\shortauthor{L. Santelli, G. Wang, R. J. A. M. Stevens, and R. Verzicco}
\title{Characterization of natural convection between spherical shells}
\author{
Luca Santelli \aff{1}
\corresp{\email{luca.santelli@gssi.it}}
Guiquan Wang\aff{2},
Richard J. A. M. Stevens\aff{2}
 \and Roberto Verzicco\aff{1,2,3},
  }

\affiliation{
\aff{1}Gran Sasso Science Institute, Viale F. Crispi 7, 67100 L'Aquila, Italy
\aff{2}Physics of Fluids Group and Twente Max Planck Center, Department of Science and Technology, Mesa+ Institute, and J. M. Burgers Center for Fluid Dynamics, University of Twente, P.O. Box 217, 7500 AE Enschede, The Netherlands
\aff{3}Dipartimento di Ingegneria Industriale, University of Rome' Tor Vergata', Via del Politecnico 1, 00133 Rome, Italy
}

\begin{document}

\maketitle

\begin{abstract}
    In this manuscript, it is analysed the onset and evolution of natural convection of an incompressible fluid between spherical shells. The shells are kept at a fixed temperature difference and aspect ratio, and the Rayleigh-Bénard convection is driven by different radial gravity profiles. The analysis has been carried out by using a finite difference scheme to solve the three-dimensional Navier-Stokes equations in spherical coordinates.
    Numerical results are compared with theoretical predictions from linear and non-linear stability analysis, and differ from the expected critical Rayleigh number $Ra_c\approx1708$ by less than $1\%$.
    
    In the range of Prandlt numbers $Pr$ studied, and for all the different gravity profiles analysed, the system presents a dependence on its starting condition and flow history. Even in the region just above the onset of convection, two stable states are observed, with qualitative and quantitative differences, and exploring higher values of $Ra$ introduces new modes and time dependency phenomena in the flow. These results are corroborated by spectral analysis. 
    
\end{abstract}

\begin{keywords}
Authors should not enter keywords on the manuscript, as these must be chosen by the author during the online submission process and will then be added during the typesetting process (see http://journals.cambridge.org/data/\linebreak[3]relatedlink/jfm-\linebreak[3]keywords.pdf for the full list)
\end{keywords}

\section{Introduction}
Natural convection in spherical domains, i.e. the convective behaviour of a fluid confined between two spherical shells and subjected to a temperature gradient, has been the focus of many studies thanks to the vast number of possible applications in different fields.
This topic has been studied experimentally \citep{bishop1966heat}, analytically \citep{mack1968natural}, and more recently numerically \citep{garg1992natural}. Many modern applications include geoscience, cosmoclimatology \citep{svensmark1997variation, svensmark2007cosmoclimatology}, as well as exploration of extraterrestrial moons \citep{feldman2012simulation} and various engineering applications. 

When considering convection between spherical shells, it should be taken into consideration that the behaviour is different from the classical Rayleigh-B\'enard convection in a planar layer configuration \citep{siggia1994high, Ahlers2009, Chilla2012}. The reasons behind this difference can be found in the geometrical asymmetry between the inner and outer sphere, the curvature of the plates and the radial dependence of buoyancy \citep{busse1970differential, spiegel1971convection,o2013comparison,Gastine2015}.


When considering a radial gravity profile, a large part of the literature is focused on internal heat source problems, mostly due to their relevance to geophysics \citep{busse1975patterns,joseph1966subcritical}. Moreover, performing direct experiments with radial gravity on the surface of the Earth is a very demanding task, due to the presence of vertical gravity. Therefore, the only experiments of this kind that have been performed were all during space missions: on the Space Shuttle \citep{hart1986space} and on the International Space Station \citep{futterer2008thermal, travnikov2003geoflow}, where a radial electrostatic potential with a profile $g(r)\propto r^{-5}$ has been used to model gravity. In the studies of the onset of convection, \citet{busse1975patterns} found qualitative differences between convective patterns of odd and even spherical harmonic order by the perturbation analyses where solutions with different wavenumber differ only quantitatively in a planar configuration. With $Ra$ up to 100 times larger than the critical value, \citet{bercovici1989three} found the convective pattern persists with the solutions by perturbation analyses. When $Ra$ is as high as $Ra>10^5$, \citet{iwase1997interpretation} showed that the axisymmetric convective patterns breakdown and flow patterns start to show time-dependent behavior, which is characterized by upwelling and downwelling thermal plumes \citep{bercovici1989three, yanagisawa2005rayleigh, Futterer2013, Gastine2015}. 
To better understand the onset of convection, it is important to consider results obtained from the linear stability analysis. Concerning spherical shells, early studies on the topic can be found on the works of \citet{chandrasekhar1961hydrodynamic} and \citet{joseph1966subcritical}. More recently, the topic has been expanded by \citet{araki1994thermal}, which used a radial gravity profile and a thin layer to perform a linear asymptotic analysis, comparing theoretical results with numerical simulations. From this work, further development has been provided by \citet{Avila2013}, which found a lower boundary for $Ra_c$ and identified some examples of most unstable modes. Finally, \citet{mannix2019weakly} included considerations for weakly nonlinear mode interactions, which might show a dependence on $Pr$ for the stable modes, under the assumption of axisymmetric spherical convection.



For what concern numerical schemes, \citet{Verzi96} have shown that, in the case of cylindrical problems, symmetries in the grid structure may cause perturbations in the fluid evolution, thus it is suggested to use models with an appropriate grid symmetry. \citet{Gastine2015} raise a similar concern, suggesting that a direct application of planar geometry models to spherical models is questionable, and direct numerical simulations in spherical shells are required. 
There exist some numerical simulations for non rotating radial gravity model with finite Prandtl number \citep{Gastine2015,tilgner1996high}, but the vast majority of them has been done with infinite Prandtl number, since this can be used to model a good approximation of the Earth mantle \citep{zebib1980infinite,bercovici1989three}.

Thus, in this manuscript we want to characterise the Rayleigh-Bénard convection between spherical shells in a non-rotating environment using a finite-difference scheme in spherical coordinates. We consider the effect of different radial gravity profiles and Prandtl number in the exploration of the flow history, giving due attention to the flow structure and wavenumber analysis.

This paper is organised as follows: the physical problem is described in section \ref{sec:numsim_rbc} together with the numerical discretisation; in section \ref{sec:stability} a linear stability analysis is presented,; in section \ref{sec:results_rbc}  are shown the behaviour of $Nu$ as a function of $Ra$ and the spectral analysis for both water and air, together with the rest of the results; a final discussion is presented in section \ref{sec:conclusion_rbc}.

\section{Numerical simulation} 
\label{sec:numsim_rbc}

\subsection{Problem description}

\begin{figure}
\begin{center}
\includegraphics[width=6cm]{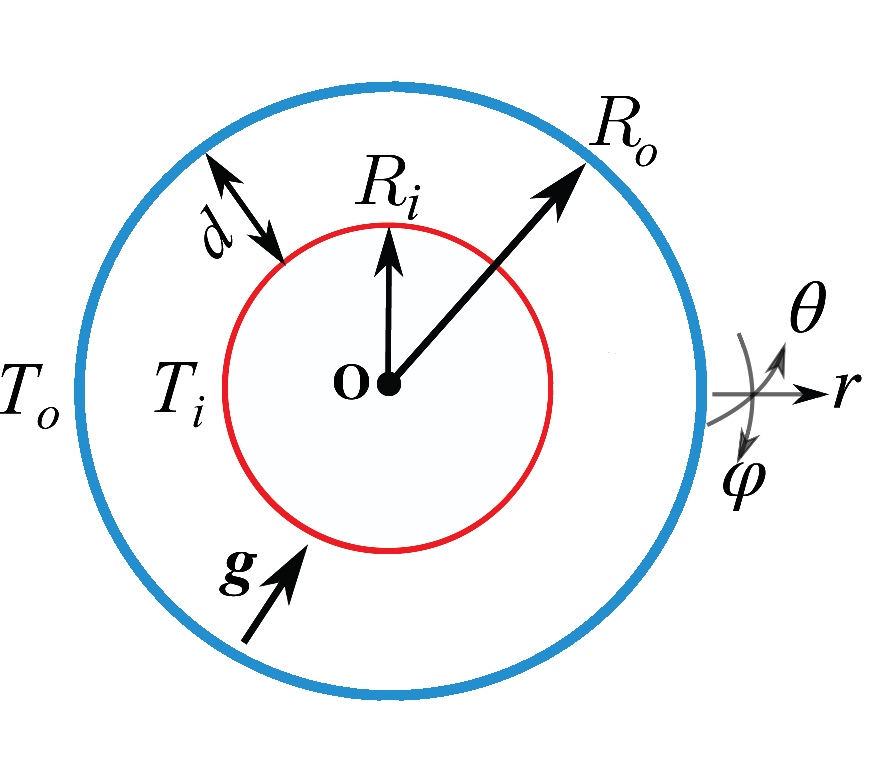}
\caption{Sketch of the configuration.}
\label{fig:fig1a}
\end{center}
\end{figure}
The configuration of the problem, a sketch of which is given in figure \ref{fig:fig1a}, consists of an inner spherical shell of radius $R_i$ and an outer concentric shell of radius $R_o$. We call $d=R_o-R_i$ the distance between the shells. The aspect ratio $\eta=R_i/R_o$ is fixed at $\eta=0.71$. The fluid is subjected to a radial gravity of the form $\vb{g}(r)=g_o \vb{g}^*(r)$, where $g_o$ is the magnitude of the gravity at the outer sphere and $\vb{g}^*(r)=\bar{g}(r)\vu{r}$ a dimensionless radius-dependent function. A fixed temperature is set for inner and outer walls, with the outer wall temperature $T_o$ being lower than the inner one $T_i$, and $\Delta T = T_i-T_o$ the temperature difference. No-slip boundary conditions are chosen.

Let $\nu$ the thermal viscosity and $\kappa$ the thermal diffusivity of the fluid, then the Prandtl number is defined as $Pr=\nu/\kappa$. 
In this paper, values of $Pr_{air}=0.71$ and $Pr_{water}=7.1$ are used; when not specified, it is assumed we are using $Pr=Pr_{air}$.
Being $\alpha$ the thermal expansion coefficient, the Rayleigh number is defined as $Ra=\frac{g_o\alpha \Delta Td^3}{\nu\kappa}$, and it will be used as the main control parameter in the following sections. 

Reynolds number is defined as $Re=Ud/\nu=\sqrt{Ra/Pr}$, where U is a free-fall velocity $U=\sqrt{g_o\alpha\Delta Td}$. The introduction of $U$ allows us to set the representative scale for length ($d$), time ($d/U$) and temperature ($\Delta T$).

The Nusselt number $Nu$ is used to measure the dimensionless heat transfer between shells, and it has been computed by direct measurement of heat flux at outer and inner shells 
\begin{equation}
\label{eq:nusselt}
Nu=\eta\pdv{\overline{ T^*}}{r}\eval_{R_i}=-\frac{1}{\eta}\pdv{\overline{T^*}}{r}\eval_{R_o},
\end{equation}
with $\overline{T^*(r)}$ being the dimensionless temperature averaged over time and surface (computed respectively at the inner and outer radius), and the equality between the two definitions is true for a (statistically) steady flow. The two different definitions of $Nu$ have been compared for all the simulations, always showing an excellent agreement.

Computing a typical diffusive time $t_d=Re/Nu$ ensures that every simulation is run for sufficient time.

Finally, the density of the system is defined as $\rho=\rho_o\alpha T$, with $\rho_o$ being the density at the outer shell. Using these quantities the problem is defined by a dimensionless Navier-Stokes equation for an incompressible viscous fluid under the Boussinesque approximation that reads as:
\begin{equation}
\label{eq:NavierStokes}
\begin{cases}
&\frac{\text{D}\vb{u}^*}{\text{D}t^*}=-\grad p^*+ T^* \vb{g}^*(r)+\sqrt{\frac{Pr}{Ra}}\laplacian \vb{u}^*\\
&\div{\vb{u}^*}=0\\
&\frac{\text{D}T^*}{\text{D}t^*}=\frac{1}{\sqrt{RaPr}}\laplacian T^*
\end{cases}
\end{equation}
with $\vb u^*$ and $p^*$ being respectively the dimensionless velocity and pressure.

\begin{figure}
\begin{center}
\includegraphics[width=0.7\textwidth]{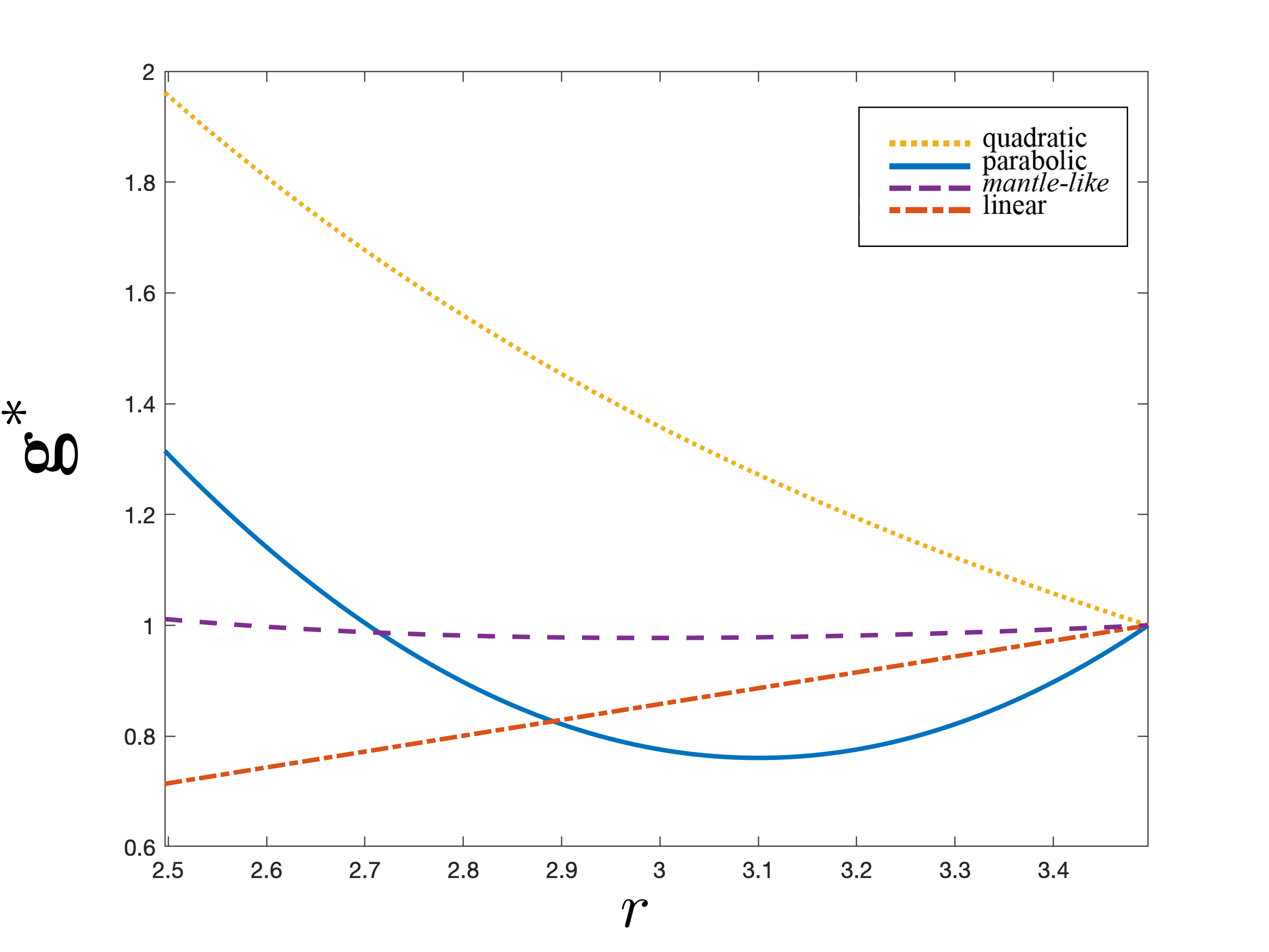}
\caption{Overview of the dimensionless gravity profiles used (summarised in table \ref{tab:gravity profiles}) as a function of radius: %
{\color{orange} \dotted }quadratic gravity; %
{\color{blue} \solid }parabolic gravity; %
{\color{violet} \dashed }\textit{mantle-like} gravity;  %
{\color{red} \chndash }linear gravity. %
}
\label{fig:gravity_profiles}
\end{center}
\end{figure}

\begin{table}
    \centering
\begin{tabular}{c||c|c|c|c|c}
\hline
Name            & Quadratic & Constant & Linear & \textit{Mantle-like} & Parabolic \\
\hline
Symbol          &$\vb g^q$  & $\vb g^c$ & $\vb g^l$ & $\vb g^m$                & $\vb g^p$\\
Equation        &$\frac{1}{r^2}$ & 1     & $r$   & $\frac{R_i^3}{r^2}(1-\lambda)+\lambda r$ & $(r-R_m)^2+\Delta R$
\end{tabular}
\caption{Different gravity profiles. $\lambda$, $R_m$ and $\Delta R$ are parameters.}
\label{tab:gravity profiles}
\end{table}

\begin{figure}
\begin{center}
\includegraphics[width=0.7\textwidth]{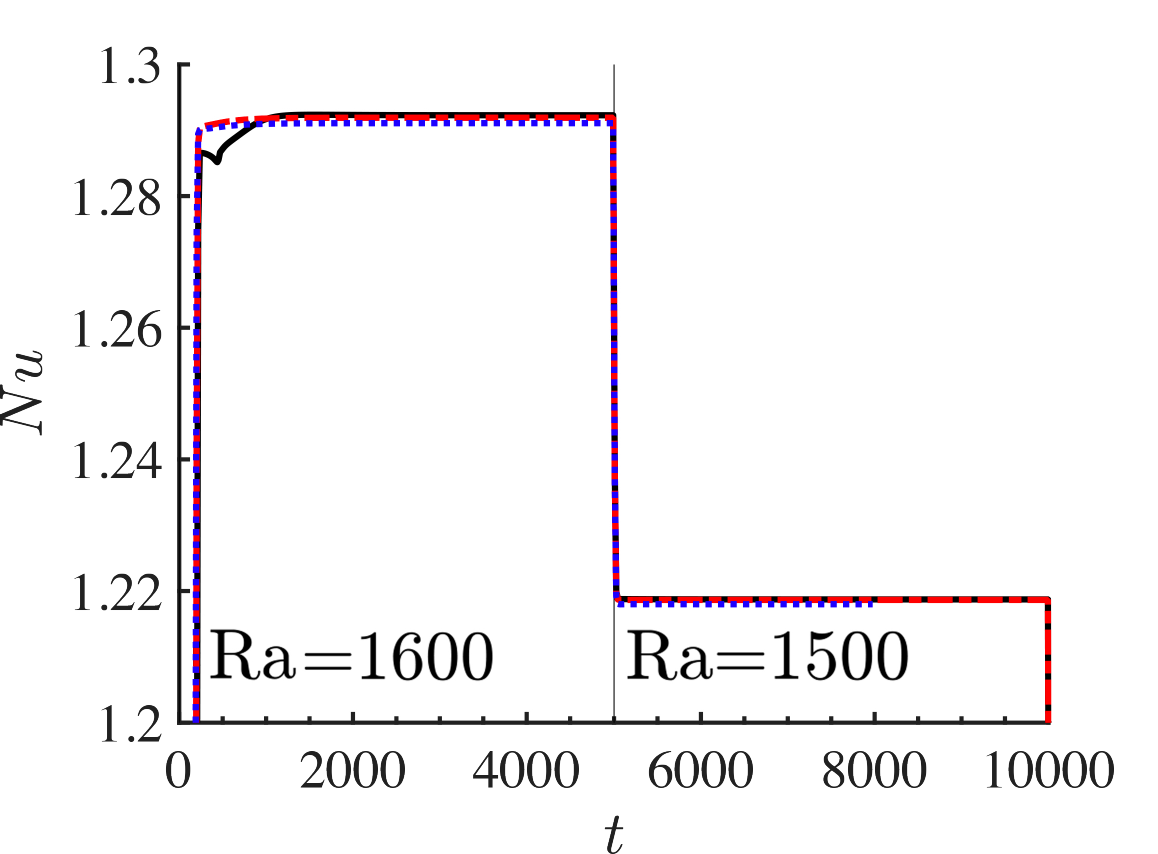}
\caption{Analysis of $Nu$ vs $t$ at different number of grid points. %
{\color{black} \solid }grid with ($N_\theta=33,N_r=35,N_\phi=35$) points; %
{\color{red} \chndash }($65,69,69$); %
{\color{blue} \dotted }($129,121,121$).  %
Only a small mismatch at the beginning of the simulation is present for the less refined grid, while the other two are almost coincident at every step. A vertical black line separate the two different $Ra$ analysed: $Ra=1600$ for $t<5000$, $Ra=1500$ after.}
\label{fig:quad_grid}
\end{center}
\end{figure}

Different gravity profiles have been used for the simulations. They are schematized, together with the associated symbols, in table \ref{tab:gravity profiles}, while their shape is shown in figure \ref{fig:gravity_profiles}. The \textit{Mantle-like} gravity profile models a situation in which there are two densities in the system: the density of the fluid between the two shells $\rho_o$ and the density inside the inner shell $\rho_i$. Their ratio $\lambda=\rho_o/\rho_i$ is the parameter describing different configurations: in the \textit{Mantle-like} gravity profile it is set to $\lambda=10/29$, while $\lambda=1$ represents the linear case, and the quadratic case is obtained as $\lambda\to 0$. The constant gravity is instead a good approximation of the situation on the surface of the Earth, and can be obtained from the mantle-like profile as $R_i\to\infty$. Finally, the parabolic profile is an artificial profile created to simulate a highly non-monotonic gravity, with $R_m=3.1$ and $\Delta R=0.5$ being parameters chosen to have a non-monotonic asymmetric (with respect to the middle radius) profile. In the following sections results will be shown for the case $\vb g= \vb g^q$, and the behaviour of other gravity profiles will be discussed alongside. 

Henceforth asterisks of dimensionless quantities are dropped in order to simplify the notation.

\subsection{Numerical setup}
\label{sec:NumericalSetup}
The Navier-Stokes equation coupled with the Boussinesque approximation has been written in spherical coordinates and discretised on a staggered spherical grid. The finite difference scheme used is second order in time and space, and exploits a change of variables to obtain trivial boundary conditions at the poles (which were the source of singularities) and the special treatment of a few discrete terms. One of the advantages of this scheme is that it allows non uniform grids in latitudinal and radial directions. More details on the method, together with a deep analysis of performances and accuracy, can be found in \citet{santelli2020finitedifference}, which extends to spherical coordinates the idea of \citet{Verzi96}.
Being the singularity at the center of the sphere outside of our analysed domain, the parallellization of the code is easy to implement. The appropriate space resolution has been chosen by running several simulations with varying grid spacing. In figure \ref{fig:quad_grid} is shown the behaviour of $Nu$ as a function of time for different grid resolutions and values of Rayleigh number: the difference between the two most refined grid shown is negligible; therefore, we can save computational time by running simulations on a grid $\qty{N_\theta=65, N_r=69, N_\phi=69}$. A maximum $CFL=0.8$ has been imposed to control the size of the time step during the convective phase, while $\Delta t_{max}=10^{-3}$ has been fixed for the linear evolution.


\begin{figure}
\begin{center}
\subfloat[\label{fig:quad_Nu-vs-Ra_inc}$Nu$ vs $t$, increasing from $Ra<Ra_c$]{\includegraphics[width=0.470\textwidth]{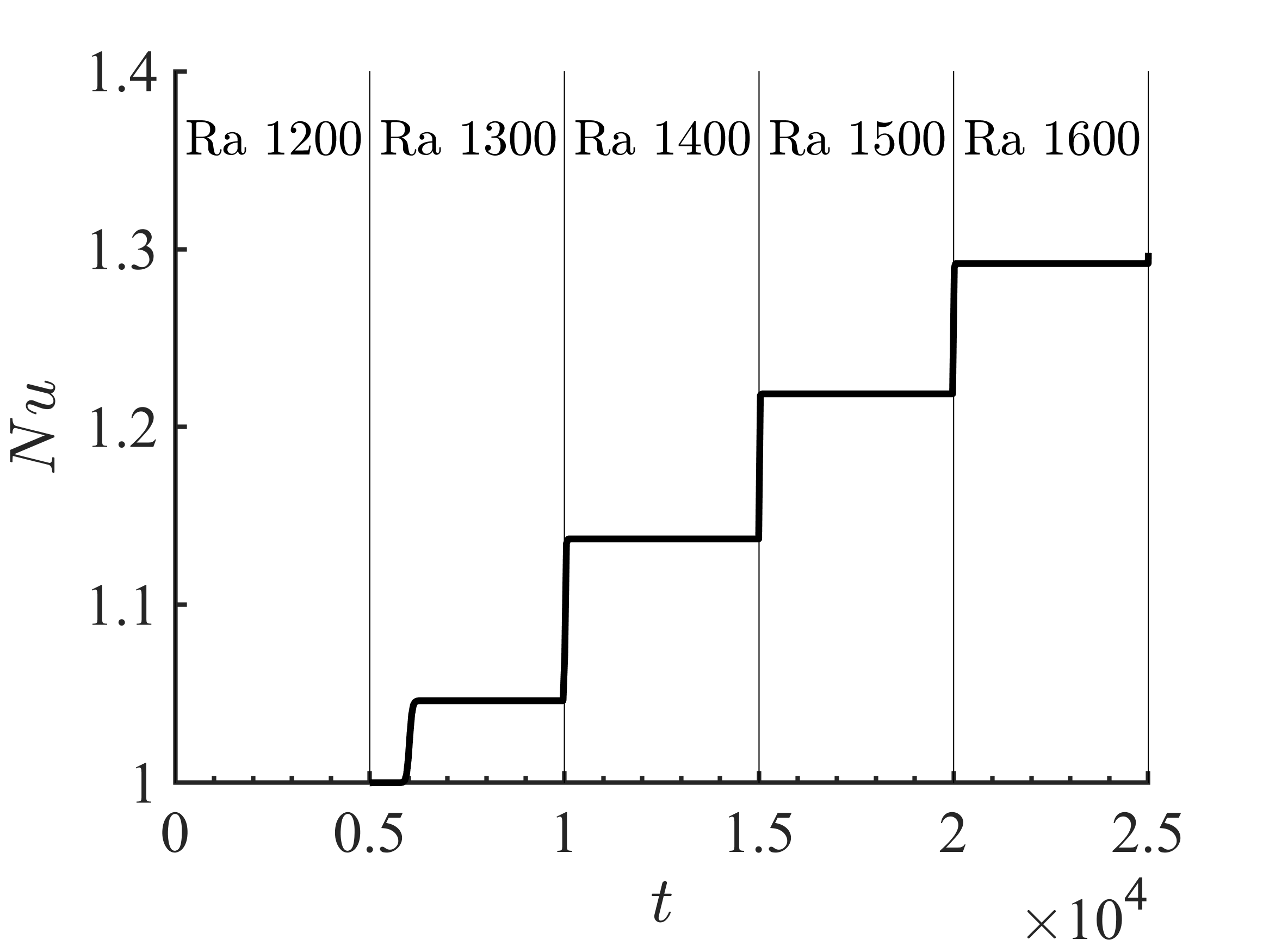}} \hfill
\subfloat[$Nu$ vs $t$, decreasing from $Ra\gg Ra_c$]{\includegraphics[width=0.470\textwidth]{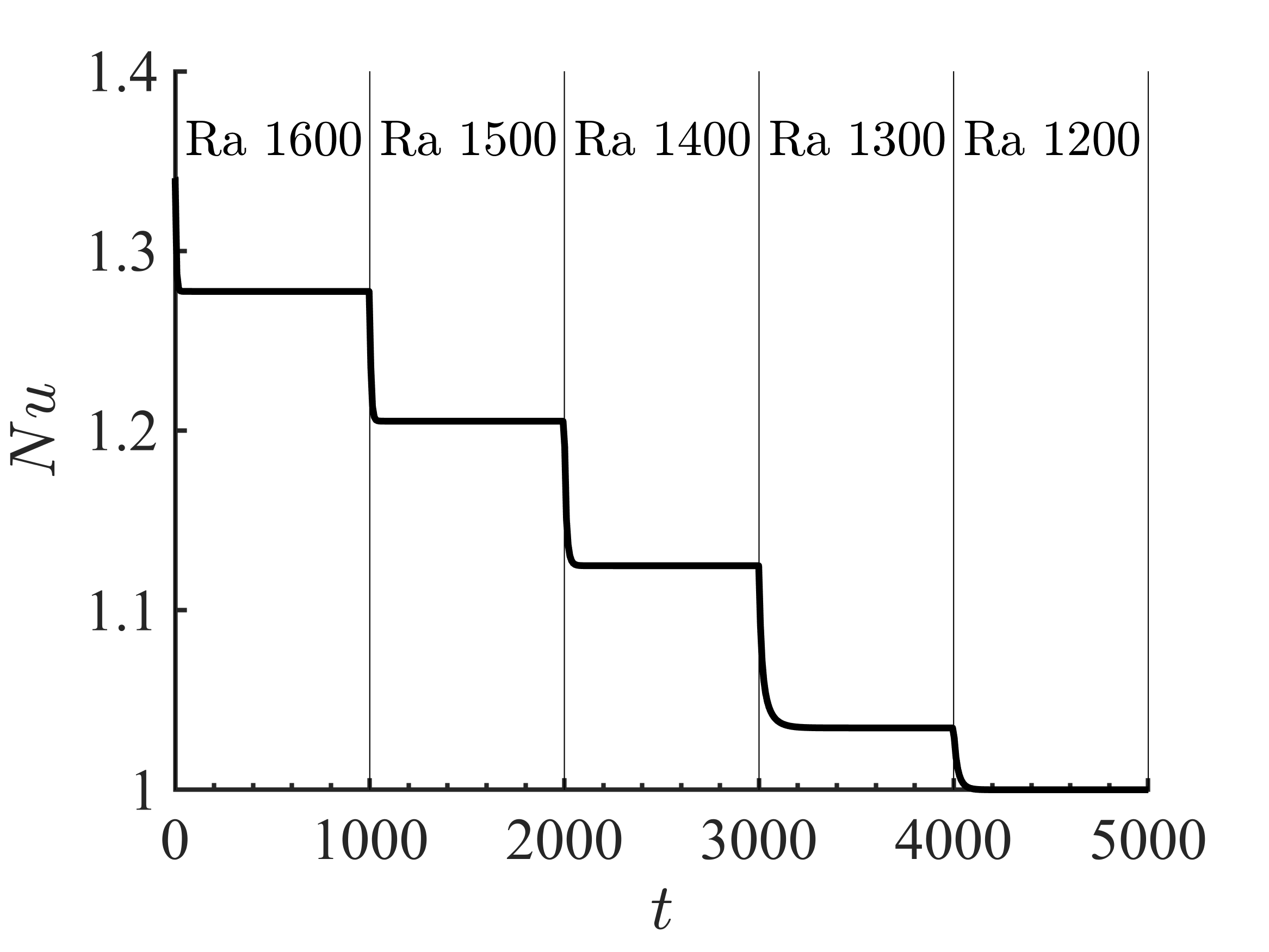}} 
\caption{Evolution of $Nu$ as function of $t$.  Vertical bars of (a) and (b) separate zones with different $Ra$.}
\label{fig:quad_Nu-vs-Ra}
\end{center}
\end{figure}

The initialization and evolution of the system varies according to the specific parameters that have been used for each simulation: 
\begin{itemize}
    \item Low $Ra$, increasing: for these simulations the flow is initialised to a rest state at $Ra<Ra_c$, with $Ra_c$ being the critical Rayleigh number for the onset of convection. It is then allowed to evolve unperturbed for a sufficiently long time in order to reach and maintain an equilibrium state. At this point, $Ra$ is increased to a new value and the equilibrium procedure is repeated. These steps are repeated until $Ra$ reached the desired value. This long-time evolution is mostly necessary for the first steps, as the absence of perturbation and the accuracy of the code drastically increase the needed time for the onset of convection.
    \item Low $Ra$, decreasing: These simulations follow the same procedure as the increasing case, but the initialization is done at $Ra\ge Ra_c$ and the subsequent values of $Ra$ are in a decreasing order. Less time per $Ra$-step is needed, since the system stabilise faster.
    \item High $Ra$: simulations done at higher values of Rayleigh number focus on the properties of the flow at a fixed value of $Ra$; therefore, after the initialization at the chosen $Ra$ from a pre-existing simulation, the fluid is left free to evolve without variations in parameters for a time sufficiently long to observe the complete behaviour.
\end{itemize}
The few cases that require a different approach will be described in their relative sections. In all the simulations, the flow needed a few time steps to reach a divergence free state, thus data are only collected after this condition has been reached.

\section{Linear stability analysis}
\label{sec:stability}
In this section, we give an overview of the work of \citet{araki1994thermal} and follow their approach to perform the linear stability analysis for this study. For this analysis (and only in this subsection), we redefine $g_0$ as the gravity at the inner radius, and perform an analysis for $R_i\gg1$. We note that the parabolic gravity profiles diverges as $R_i^2$ when $R_i\to\infty$, thus it is ill-suited for this analysis and should not be considered here. On the other hand, the constant gravity profile is included here as the limit of the mantle-like profile for $R_i\to\infty$.

We observe that equation \eqref{eq:NavierStokes} can be rewritten as 
\begin{equation}
\label{eq:NavierStokesVTHETA}
\begin{cases}
&\laplacian\qty(\laplacian -\frac{1}{Pr}\pdv{t})\vb V = Ra\frac{g(r)}{R_i}\qty(-\frac{1}{\sin\varphi}\pdv{\varphi}\sin\varphi\pdv{\varphi}-\frac{1}{\sin^2\varphi}\pdv[2]{\varphi})\vb \Theta\\
&\qty(\laplacian - \pdv{t})\vb\Theta = -\frac{1}{r}\pdv{T}{r}\vb V R_i
\end{cases}
\end{equation}
where $\vb V=r\vb u/R_i$ and $\vb \Theta$ is the temperature fluctuation around T(r). If the fluid is stationary, spherical symmetry holds and we can represent eigenmodes of equation \eqref{eq:NavierStokesVTHETA} in terms of spherical harmonics $\vb Y_l^m(\theta, \varphi)$, and only one eigenvalue for each $l$ exist, with a degeneracy of $2l+1$ \citep{sattinger1979group}. This allow us to write 
\begin{equation}
    \mqty(\vb V\\ \vb \Theta) = \mqty(\xi_1(r,t)\\ \xi_2(r,t))\vb Y_l^m(\theta, \varphi)
\end{equation}
and focus on the equations for $\xi_1$ and $\xi_2$. Then, assuming a time dependence of the fluctuations as $\exp(\sigma t)$, the equations for $\xi_1$ and $\xi_2$ can be written as 
\begin{equation}
    \qty[\mqty(L_{22}L_{22} & L_{12} \\ L_{21} & L_{22}) - \sigma \mqty(L_{22}/Pr & 0 \\ 0 & 1)] \mqty(\xi_1(r,t)\\ \xi_2(r,t)) = \mqty(0 \\ 0) 
\end{equation}
where
\begin{align*}
    &L_{22} = \frac{1}{r^2}\pdv{r}r^2\pdv{r}-\frac{l(l+1)}{r^2}\\
    &L_{12} = \frac{l(l+1)}{R_i^2}\frac{\bar{g}(r)}{r}\\
    &L_{21} = \frac{R_i^2R_o}{r^3}\\
    &\text{and boundary conditions }\xi_1 = \pdv{\xi_1}{r} = \xi_2 = 0 \qq{at} r=R_i, R_o \qq{.}
\end{align*}

Given that the most unstable mode number $l$ diverges as $R\to\infty$, a normalised wavenumber $k=l/R$ is introduced for the analysis at $R\gg1$. Expanding $Ra(\epsilon,k)$ in powers of $\epsilon=1/R$ around $\vb c=(0,k^{(0)})$, with $k=k^{(0)}+\epsilon k^{(1)}$ and $\sigma=0$, yields
\begin{equation}
    Ra(\epsilon,k) = Ra(\vb c) + \qty(\pdv{Ra}{\epsilon}\vb{}(\vb c) +\pdv{Ra}{k}\vb{}(\vb c)k^{(1)}) \epsilon + \order{\epsilon^2}.
\end{equation}
The last term of the right-hand side is $0$ because we assume that the critical Rayleigh number satisfies $\pdv{Ra}{k}(1/R,k)=0$ at $k=k_c(0)=k^{(0)}$. 
At this point, we can write $Ra=Ra^{(0)}+\epsilon Ra^{(1)}$, obtaining a zero-order critical value of $Ra^{(0)}_c=1707.8$ and a first order correction of $Ra^{(1)}_c  = \qty(1-3/2 \lambda) Ra^{(0)}_c$. Thus, the critical $Ra$ computed up to $\order{1/R_i}$ for non parabolic gravity profiles is 
\begin{equation}
    \label{eq:critical_Ra}
    Ra_c = Ra^{(0)}_c\qty[1+\qty(1-\frac{3\lambda}{2})\frac{1}{R_i}],
\end{equation}
 with the corrective term $1/R_i$ vanishing for the constant gravity profile by construction.
According to \citet{araki1994thermal}, using a different point $x=r-R_i$ to compute the dimensionless gravity $g(x) = g_0\qty( (1-\lambda)\qty(\frac{R_i}{R_i+x})^2+\lambda \frac{R_i+x}{R_i}) $ modifies $Ra_c$ as
\begin{equation}
\label{eq:critical_Ra-x}
    Ra_c(x) = Ra^{(0)}_c\qty[1+\qty(1-\frac{3\lambda}{2})\frac{1-2x}{R_i}],
\end{equation}
implying that for $x=0.5$ the first order correction disappears. However, as we note in the following sections, we believe that a more accurate representation can be given by the introduction of an effective Rayleigh number $Ra^{e}$ defined as
\begin{equation}
\label{eq:effectiveRa}
    Ra^{e}=\frac{1}{d}\int_{R_{i}}^{R_o} \dd x Ra(x) \equiv \frac{1}{d}\int_{R_{i}}^{R_o} \dd x Ra \frac{g(x)}{g_o} ,
\end{equation}
which clearly preserves the property of vanishing first order correction if applied to equation \eqref{eq:critical_Ra-x}, while at the same time displaying a better agreement with the data.

\section{Results}

\label{sec:results_rbc}

\subsection{Onset of convection}

\begin{figure}
\begin{center}
\includegraphics[width=0.70\textwidth]{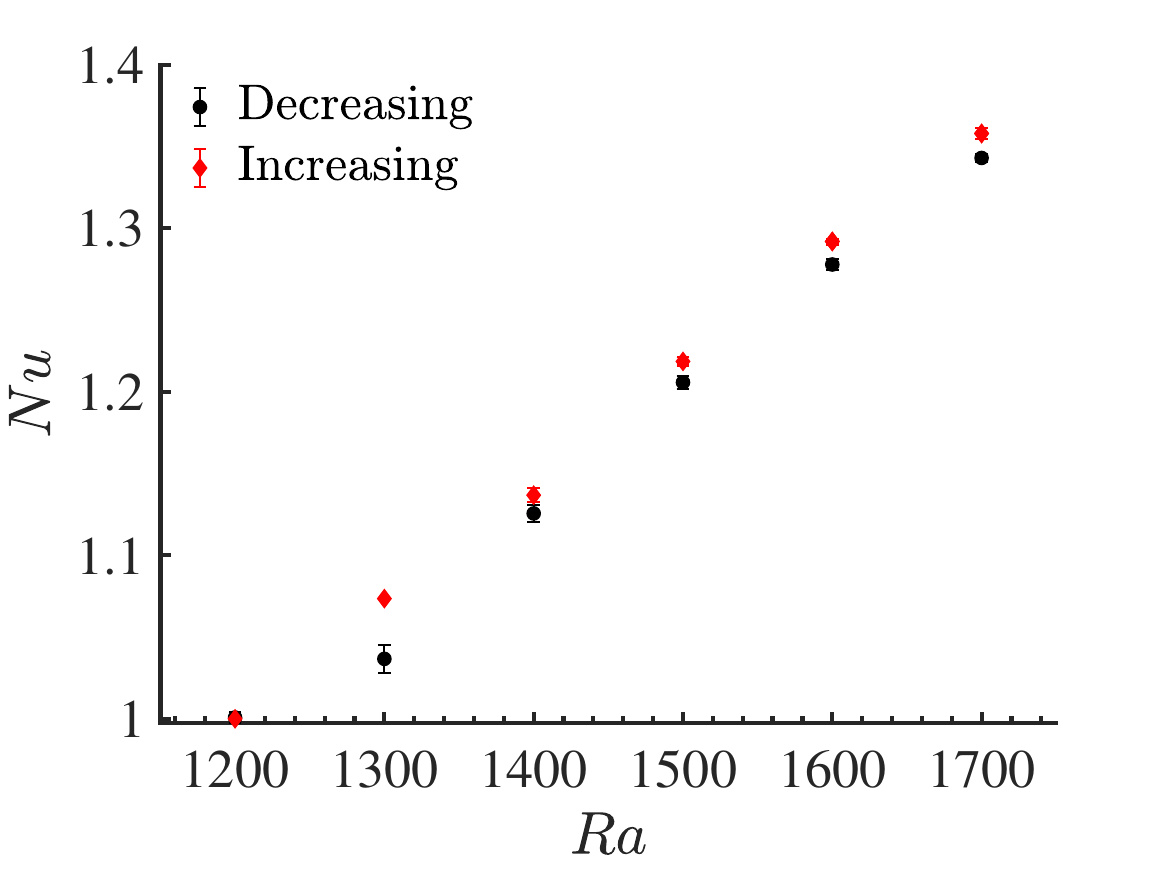}
\caption{Evolution of $Nu$ as function of $Ra$.   {\color{red} $\blacklozenge$} red diamonds for increasing-$Ra$, $\bullet$ black circles for decreasing-$Ra$. Error bars shown when bigger than the symbols.}
\label{fig:quad_Nu-vs-Ra_both} 
\end{center}
\end{figure}

The behaviour of $Nu$ as a function of $Ra$ and $t$ varies accordingly to the range of $Ra$ analysed. For $Ra$ lower then the critical value for the onset of convection, $Nu=1$. An interval of Rayleigh number around $Ra_c$ is shown in figure \ref{fig:quad_Nu-vs-Ra} for the quadratic gravity profile $g^q$. As described in section \ref{sec:NumericalSetup}, in the increasing case we set initial conditions at $Ra\le1200$, then the fluid is left free to evolve for a long enough time ($5\times 10^3$ time units for figure \ref{fig:quad_Nu-vs-Ra_inc}) until $Ra$ is updated according to $Ra_{new}=Ra_{old}+\Delta Ra$; the decreasing case starts from a much higher $Ra$ and it is then update following  $Ra_{new}=Ra_{old}- \Delta Ra$, with $\Delta Ra$ being a parameter used to refine the precision of the analysis (for figure \ref{fig:quad_Nu-vs-Ra}, $\Delta Ra=100$). For this range of values of $Ra$, $Nu(\vb r)$ is not time dependent at any fixed value of the Rayleigh number; therefore, as shown in figure \ref{fig:quad_Nu-vs-Ra_both}, the mean value of it has been computed as in equation \eqref{eq:nusselt} to compare the increasing- and decreasing- evolutions. From the figure it is immediate to notice that the two approaches yield different results. Some details on this behaviour will be given in the following paragraphs, and a deeper analysis on the phenomenon is carried out in section \ref{sec:hysteresis_rbc}. 

\begin{table}
    \centering
    \begin{tabular}{c||c|c|c|c|c}
        \hline
    Gravity & Quadratic & Linear & Constant & Mantle-like & Parabolic  \\
    \hline
  $Ra_c$      & $1240 \pm 1\%$      & $2020 \pm 1\%$   &  $1730 \pm 1\%$    & $1610 \pm 1\%$    & $2110 \pm 1\%$ \\
  $Ra_{c}^e$& $1737 \pm 1\%$      & $1735 \pm 1\%$   &  $1730 \pm 1\%$    & $1739 \pm 1\%$    & $1907 \pm 1\%$
    \end{tabular}
    \caption{Critical Rayleigh $Ra_c$ and effective critical Rayleigh $Ra_{c}^e$ for different gravity profiles. $Ra_c^e$ coincides for all the profiles except the parabolic.}
    \label{tab:critical_ra}
\end{table}

\begin{figure}
\begin{center}
\subfloat[$T$ profile at $Ra=1200$]{\includegraphics[width=0.45\textwidth]{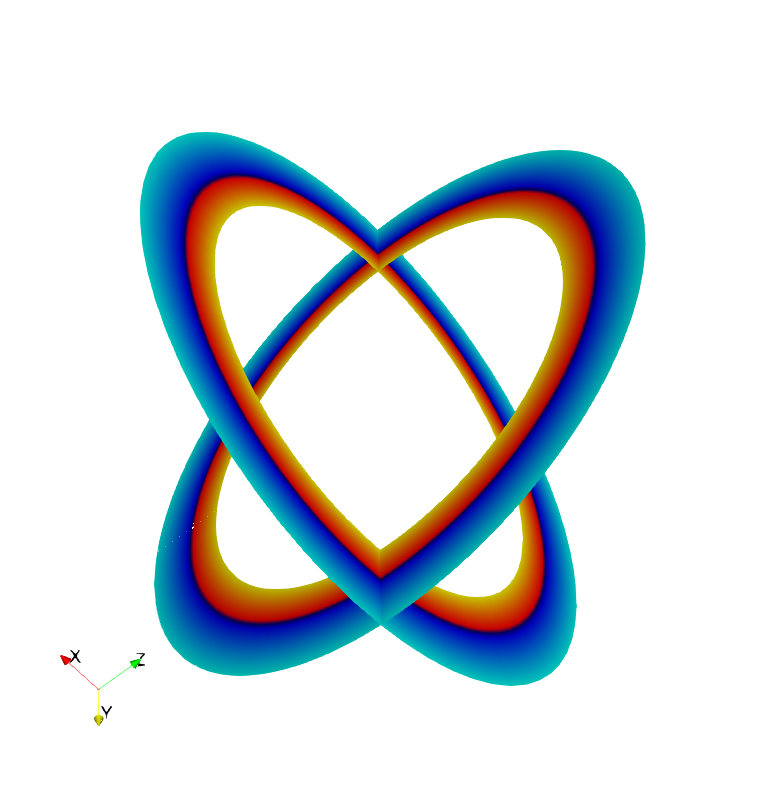}} \hfill
\subfloat[Spectrum of $Ra=1200$\label{fig:profile_very_low_Ra}]{\includegraphics[width=0.5\textwidth]{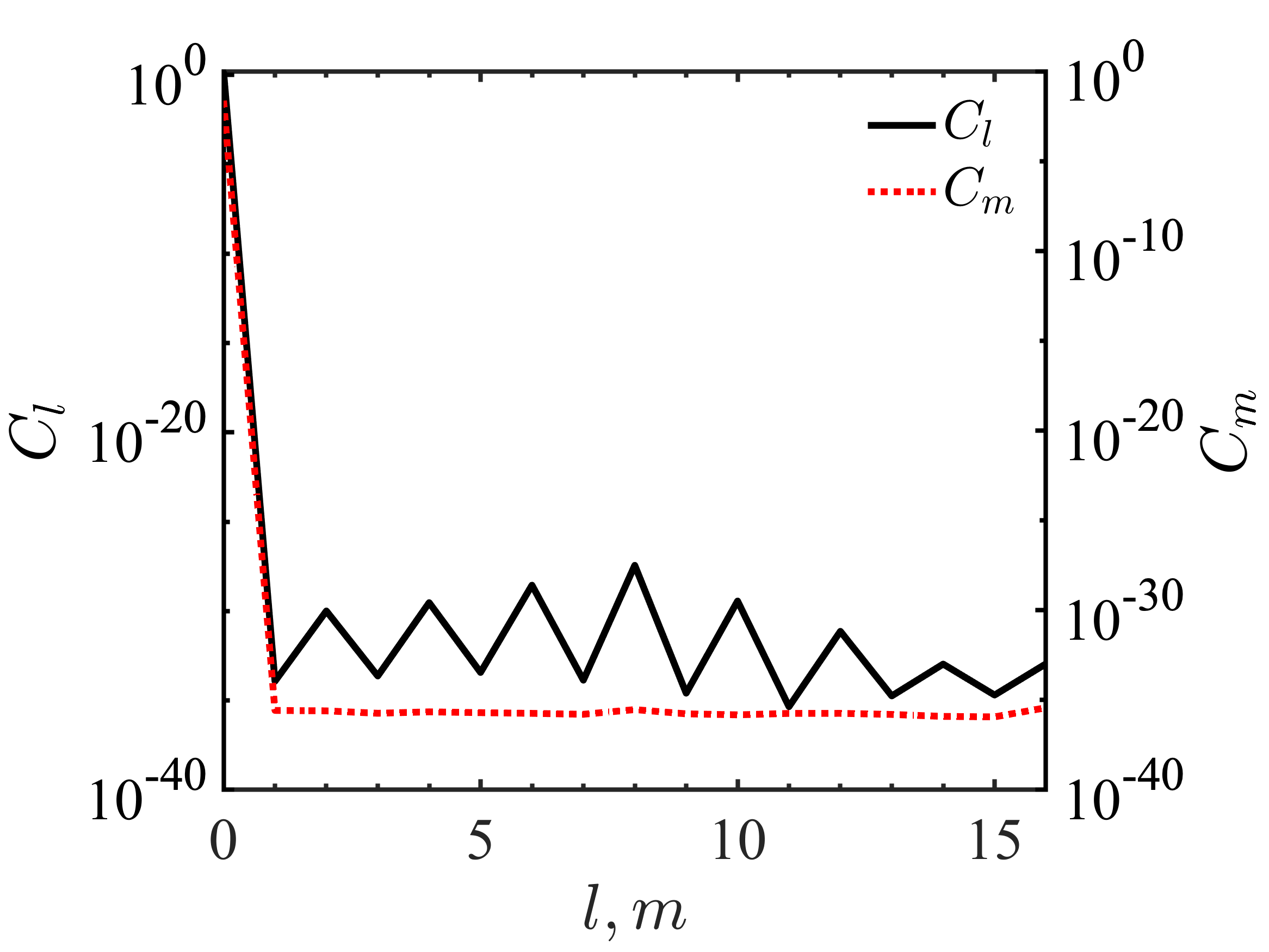}} \\[-4ex]
\subfloat[$T$ profile at increasing $Ra=1600$]{\includegraphics[width=0.45\textwidth]{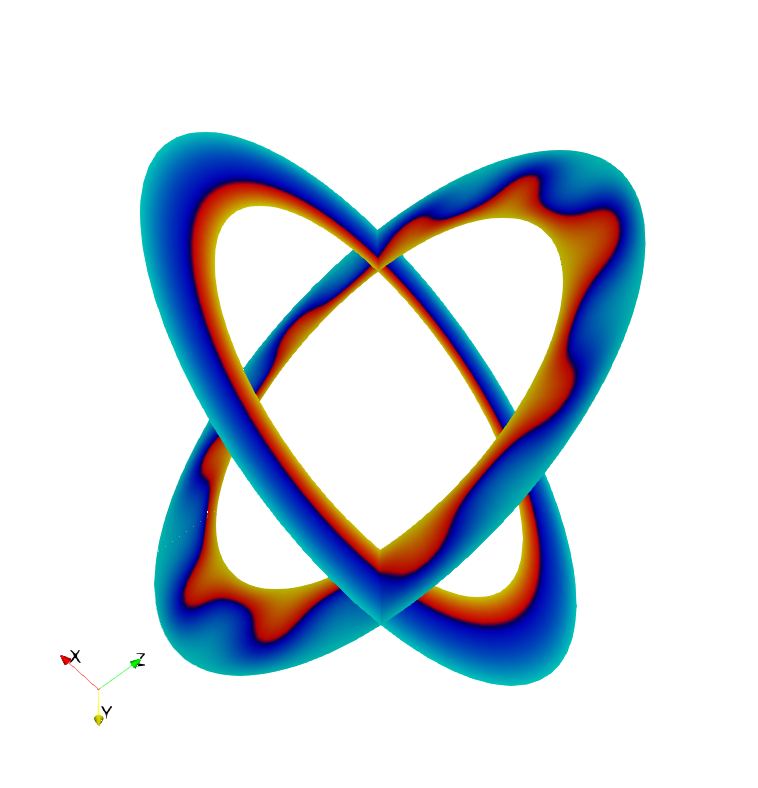}} \hfill
\subfloat[Spectrum of $Ra=1600$, increasing]{\includegraphics[width=0.5\textwidth]{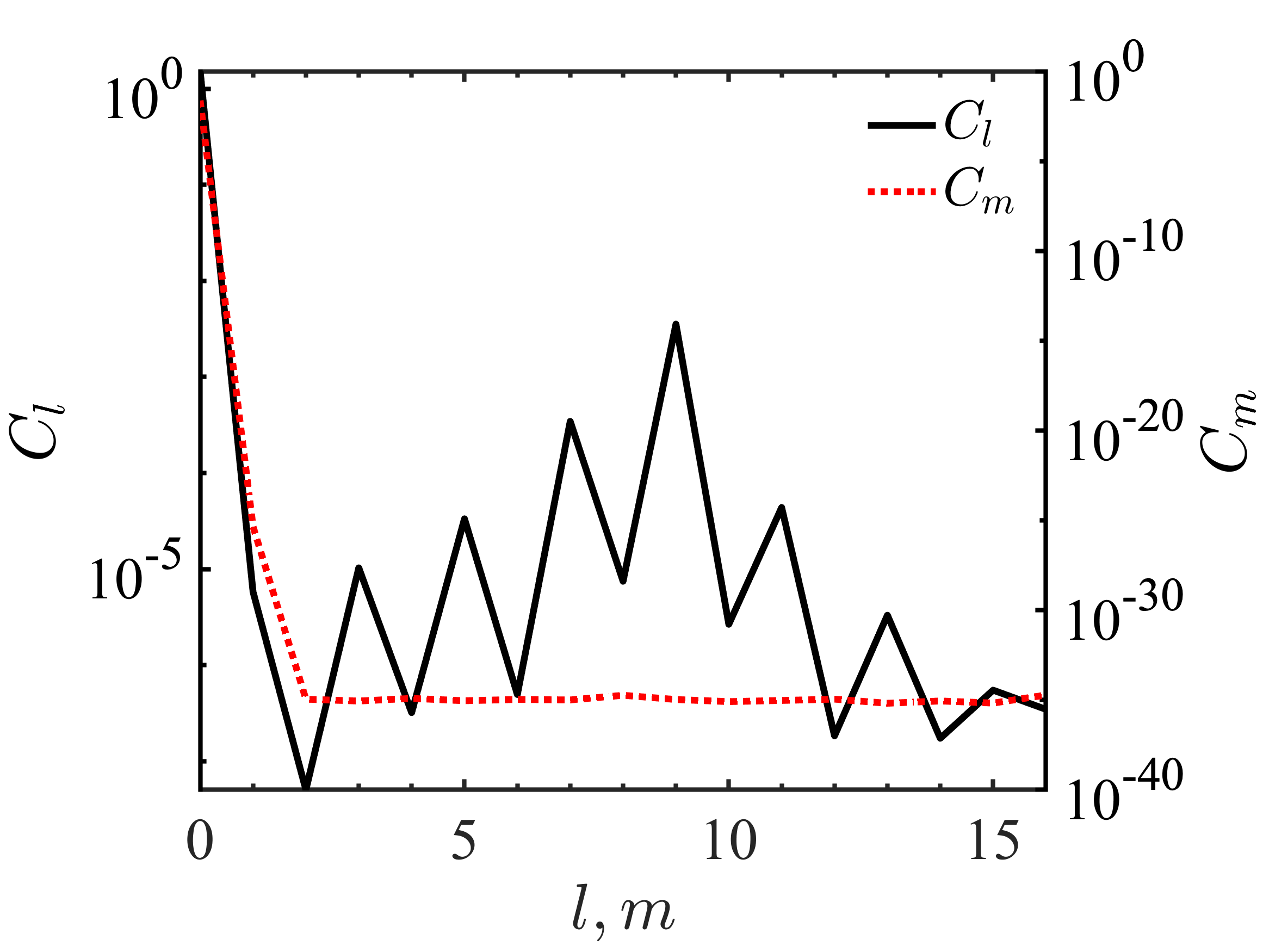}} \\[-4ex]
\subfloat[$T$ profile at decreasing $Ra=1600$]{\includegraphics[width=0.45\textwidth]{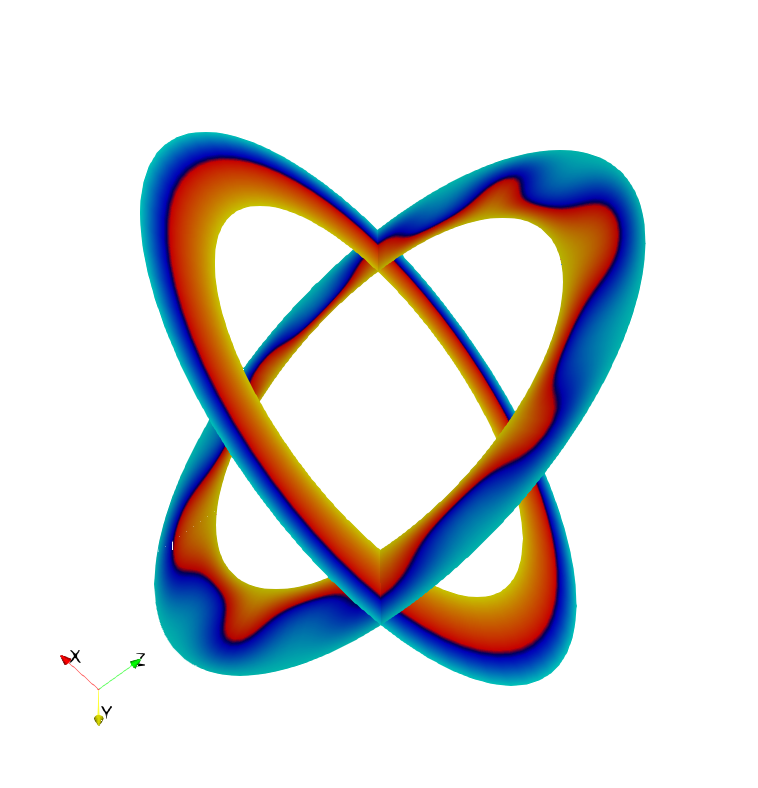}} \hfill
\subfloat[Spectrum of $Ra=1600$, decreasing]{\includegraphics[width=0.5\textwidth]{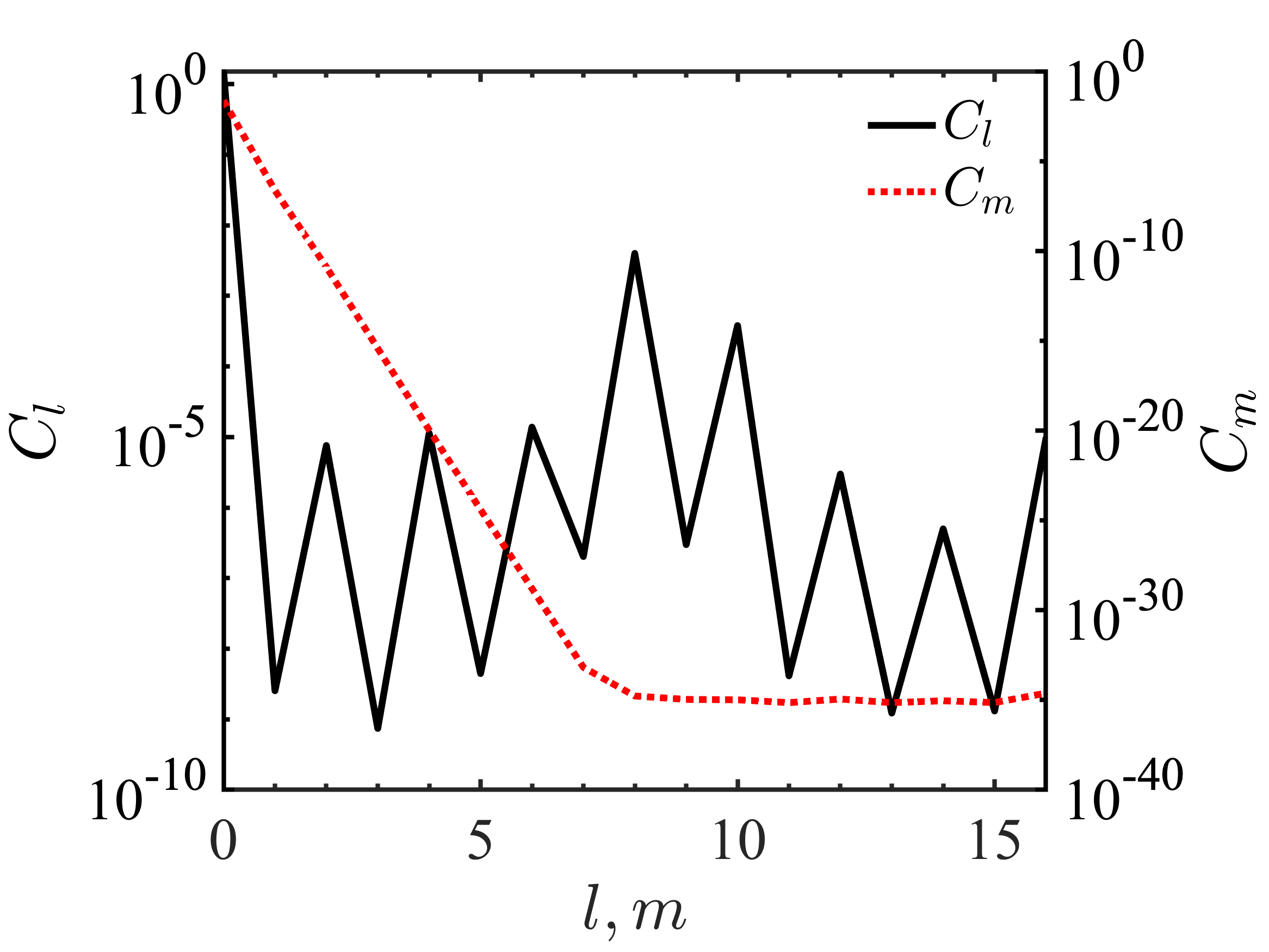}\label{subfig:try}}
\caption{Temperature profile (a,c,e) and spherical harmonic spectrum (b,d,f) for quadratic gravity in various situations. Temperature ranges from yellow (hotter) to blue (colder). For spectra, 
{\color{black} \solid }for $C_l$ and reference scale on the left; %
{\color{red} \dotted }for $C_m$ and reference scale on the right.  %
}
\label{fig:spectrum_convection}
\end{center}
\end{figure}

The critical value for the onset of convection is identified when $Nu(Ra)$ becomes greater than one,  and it is $Ra_c^q\approx 1240$ for quadratic gravity. The same study has been performed for the other gravity profiles: results from this analysis are showed in table \ref{tab:critical_ra} and are in good agreement with results from \citet{araki1994thermal}; however, for the quadratic and mantle-like gravity profiles a difference of up to $10\%$ has been observed with the expected value computed as in equation \eqref{eq:critical_Ra-x}. This is to be expected, as the stability analysis is neglecting higher order corrections that, for our values of $R_i$, can be of order of magnitude of $0.1$. Therefore, as anticipated in section \ref{sec:stability}, an effective $Ra^e$ -as defined in equation \eqref{eq:effectiveRa}- is introduced with the goal of averaging over the higher order terms.
Table \ref{tab:critical_ra} shows the importance of the effective Rayleigh number by highlighting how $Ra_c^e$ is very similar for almost all the cases analysed, differing from the theoretical critical value by less than $1\%$, with the only exception of the parabolic case, which was not covered by the previous study. We conclude this analysis by noticing that the critical $Ra$ for the onset of convection does not show any dependence on starting condition.

An important tool to understand the flow structure is represented by the analysis of the temperature profile and the corresponding spectral analysis. Both the temperature and spectral profile are shown in figure \ref{fig:spectrum_convection} for some values of $Ra$ around the onset of convection for quadratic gravity.
Details on the spectral analysis can be found in appendix \ref{app:spectrum}, here we just highlight that our focus is on the non--zero value of the degree for which the relative coefficient is higher, hereby named \textit{main-degree}. As expected, in the pure conductive case at $Ra<Ra_c$, the spectrum is almost zero for $l > 0$ and $m>0$. The corresponding temperature profile shows an unperturbed flow status. 
When convection is reached by increasing $Ra$ over $Ra_c$, the system enters in a new state, hereby named state $\mathcal{S}_9$: the main-degree is $9$ (and odd numbers dominate the $l$ spectrum), and the temperature profile has a 9-pointed shape. Analysis of the same $Ra$ reached from the decreasing case shows how the temperature profile has a 8-pointed shape, and the spectral analysis confirms that the main-degree is $8$. This state is identified as state $\mathcal{S}_8$. As we are observing a stationary condition, degeneracy for the eigenvalues is to be expected, and indeed from the data $C_m$ is negligible for any $m>0$.

\subsection{Non-stationary convection}

\label{sec:Non-stationary_convection_rbc}

\begin{figure}
\begin{center}
\subfloat[$Nu$ vs $t$ at $Ra=1700$]{\includegraphics[width=0.45\textwidth]{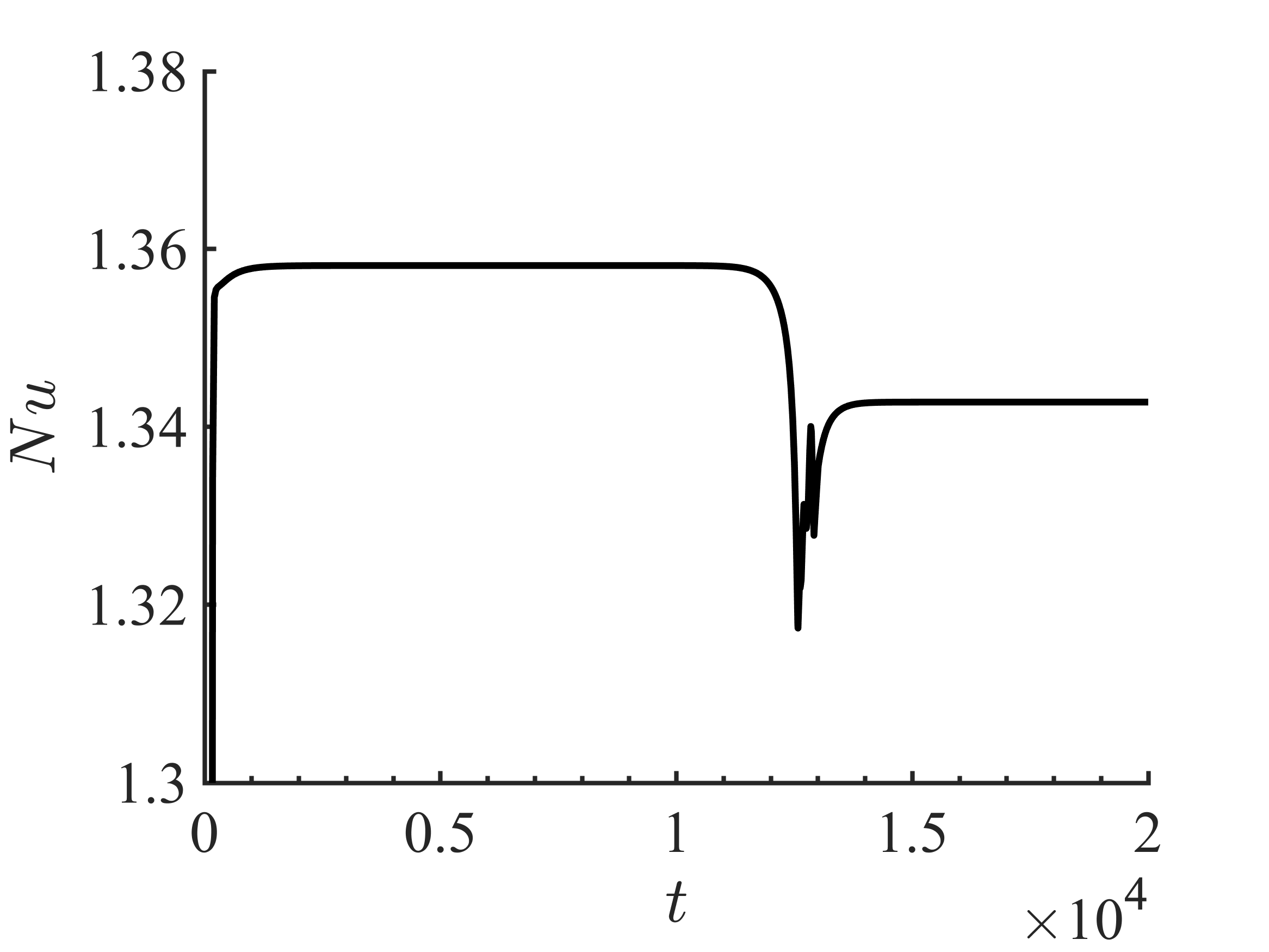}} \hfill
\subfloat[$T$ profile at $t<t_s$]{\includegraphics[width=0.25\textwidth]{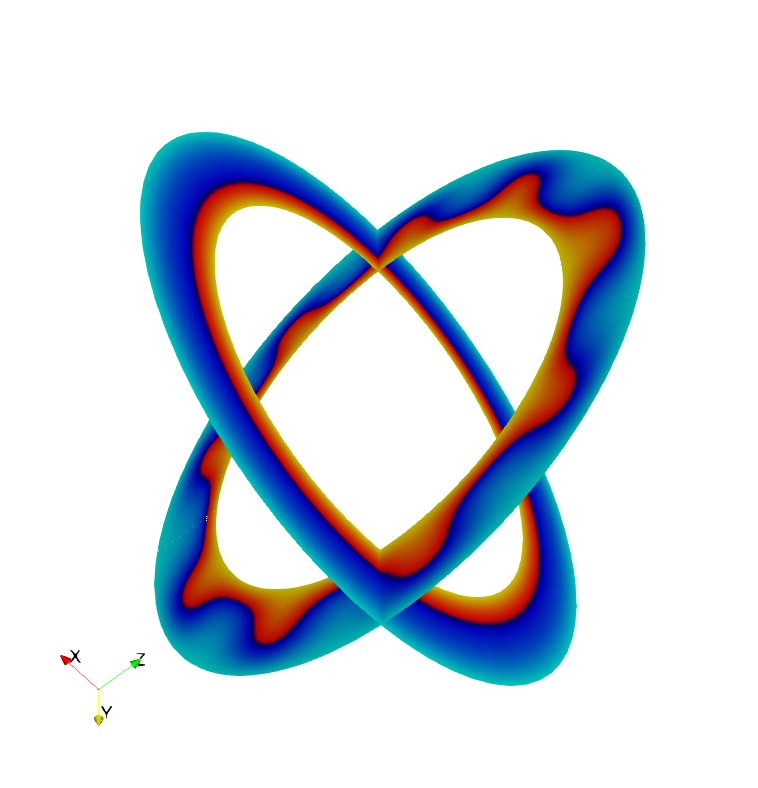}} \hfill
\subfloat[$T$ profile at $t>t_s$]{\includegraphics[width=0.25\textwidth]{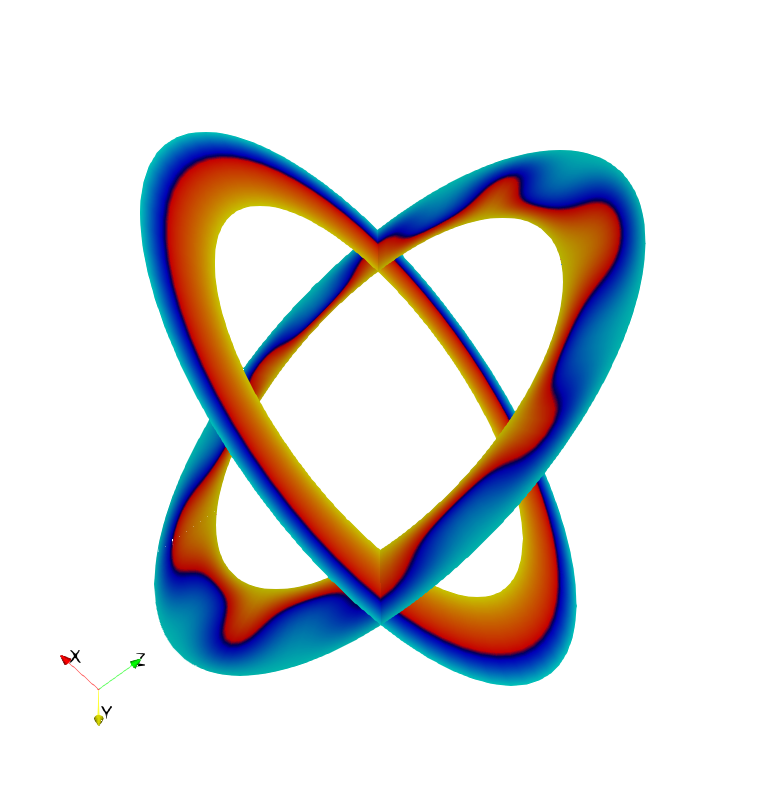}\label{fig:temperature_Ra-1700_8-subfig}} \\
\subfloat[Spectrum at $t<t_s$]{\includegraphics[width=0.45\textwidth]{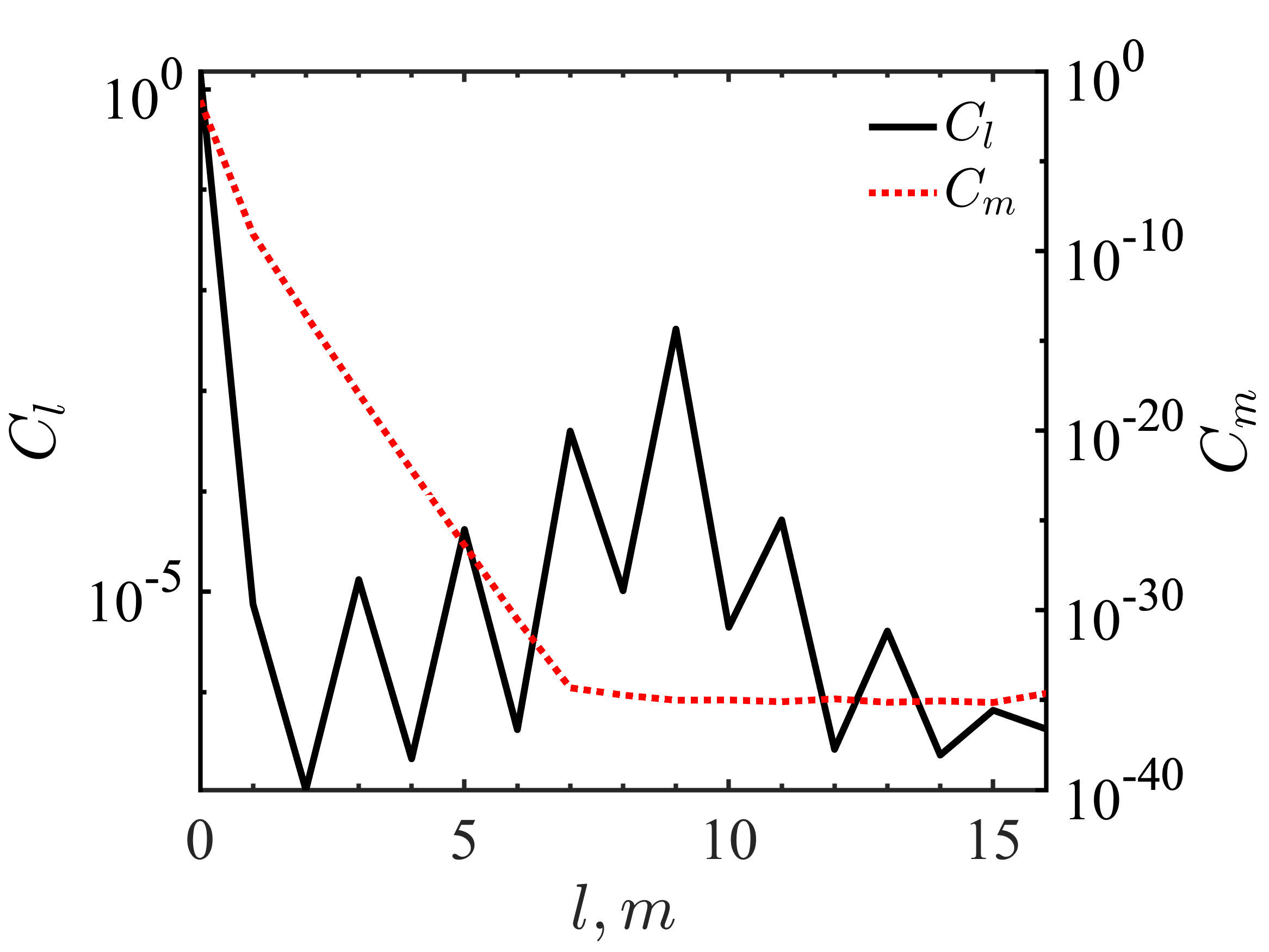}} \hfill
\subfloat[Spectrum at $t>t_s$]{\includegraphics[width=0.45\textwidth]{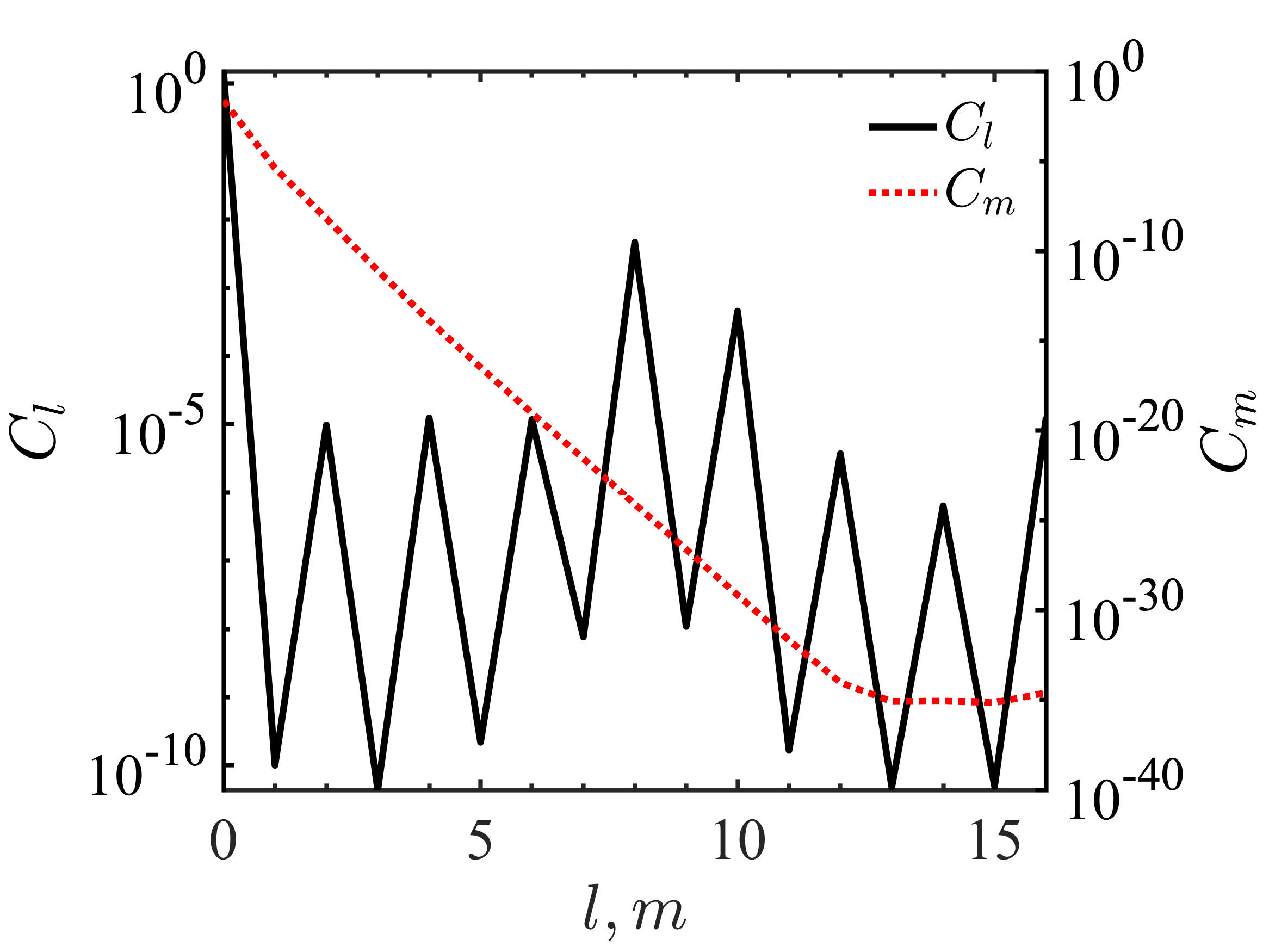}\label{fig:spectrum_Ra-1700_8-subfig}}
\caption{Analysis of quadratic gravity at $Ra_1=1700$. In figure (a) the time needed for stabilising is approximately $t_s\approx 1.2\times10^4$. 
Legend for temperature of figures (b,c) and of spectra (d,e) as in figure \ref{fig:spectrum_convection}.
}
\label{fig:Ra_1_full}
\end{center}
\end{figure}

When $Ra$ is further increased, a non-stationary behaviour appears. We can start by identifying two different regions, characterized by increasing values in the Rayleigh number: $Ra_1$ and $Ra_2$. 
We can define a region I for $Ra_1\le Ra <Ra_2$ 
and a region II for $Ra\ge Ra_2$. In region I, the system first reaches a meta stable state  $\mathcal{S}_9'$; then, after a stabilising time $t_s$ inversely proportional to $Ra$, it moves to a stable state  $\mathcal{S}_8'$.  
In figure \ref{fig:Ra_1_full} we can observe the profile of Nusselt number as a function of time for a simulation with quadratic gravity at $Ra=1700$; it is immediate to identify $t_s\approx1.2\times10^4$. In the figure, the spectral analysis and temperature profile of this system before and after $t_s$ are shown as well. As in the previous section, we notice how state  $\mathcal{S}_9'$ has a main-degree $9$, while state  $\mathcal{S}_8'$ has a main-degree $8$. To verify the relationship between  $\mathcal{S}_9$, $\mathcal{S}_9'$ and $\mathcal{S}_8$, $\mathcal{S}_8'$, we analysed the evolution of a system with initial conditions in  $\mathcal{S}_9'$ (or  $\mathcal{S}_8'$), and we decrease $Ra$ to $\tilde{Ra}<Ra_1$: the new obtained state is fully equivalent to  $\mathcal{S}_9$ (or  $\mathcal{S}_8$), thus we can identify $\mathcal{S}_9'\equiv\mathcal{S}_9$ and $\mathcal{S}_8'\equiv\mathcal{S}_8$.
The stability of $\mathcal{S}_8$ has been verified by taking a system with starting condition $\mathcal{S}_8$ and let it evolve at a different $Ra$: as long as $Ra<Ra_2$, the system will remain in state $\mathcal{S}_8$.

\begin{table}
    \centering
    \begin{tabular}{c||c|c|c|c|c}
        \hline
    Gravity & Quadratic & Linear & Constant & Mantle-like & Parabolic  \\
    \hline
    $Ra_1$      & $1650 \pm 5\%$      & $2700 \pm 5\%$   &  $2350 \pm 5\%$    & $2150 \pm 5\%$    & $2900 \pm 5\%$ \\
    $Ra_{1}^e$ & $2311 \pm 5\%$      & $2313 \pm 5\%$   &  $2350 \pm 5\%$    & $2321 \pm 5\%$    & $2621 \pm 5\%$ \\
    $Ra_2$      & $2100 \pm 5\%$     & $3400 \pm 5\%$  &  $2900 \pm 5\%$   & $2700 \pm 5\%$   & $3600 \pm 5\%$ \\
    $Ra_{2}^e$ & $2941 \pm 5\%$     & $2914 \pm 5\%$   &  $2900 \pm 5\%$   & $2915 \pm 5\%$   & $3254 \pm 5\%$
        
    \end{tabular}
    \caption{$Ra_1$, $Ra_2$ and effective value $Ra_{1}^e$, $Ra_{2}^e$ for different gravity profiles. $Ra_{1,2}^e$ coincides for all the profiles except the parabolic.}
    \label{tab:ra_1and2}
\end{table}

Given that $t_s$ increases for lower values of $Ra$, the exact value of $Ra_1$ is hard to identify. Currently, our simulations last at least for $2\times 10^5$ time units, approximately fifteen times more than the $t_s$ identified for the current $Ra_1$. Our results for both $Ra_1$ and $Ra_2$ are schematized in table \ref{tab:ra_1and2} and show the same behaviour of previous cases: results are mostly coherent once effective value has been computed, with the parabolic case being an exception (as it was in the previous analysis). 

Summarising, region I can be identified by the presence of a meta-stable flow which, after a stabilising time $t_s$ has passed, evolves into a stable time--independent behaviour\footnote{assuming it never touches region II during its evolution}.

\begin{figure}
\begin{center}
\subfloat[$Nu$ vs $t$ at $Ra=2100$. Inset: zoom on the periodic behaviour.]{\includegraphics[width=0.5\textwidth]{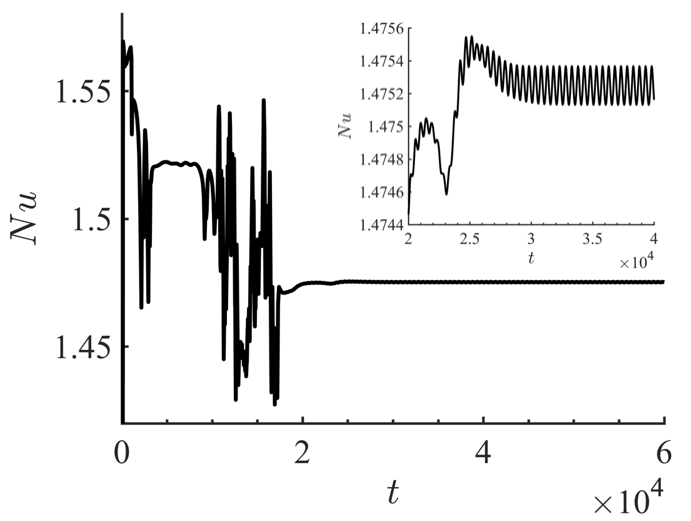}} \hfill
\subfloat[Instantaneous $T$ profile during periodic behaviour.]{\includegraphics[width=0.4\textwidth]{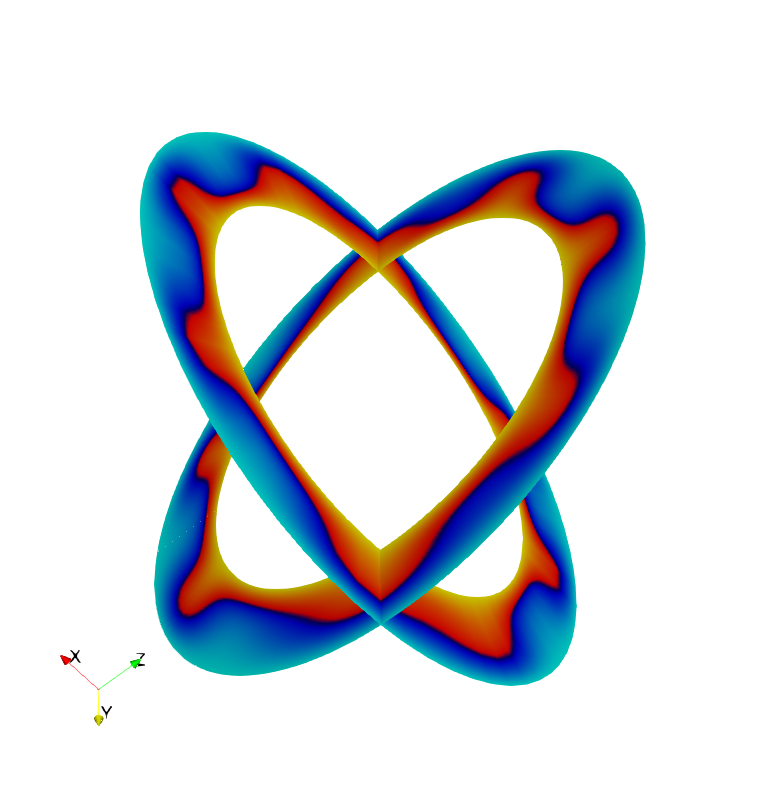}\label{fig:temperature-Ra-2100_2-subfig}} \\[-2ex]
\subfloat[Intensity $s$ vs frequency $f$. Main peak at $f\approx0.25$.]{\includegraphics[width=0.5\textwidth]{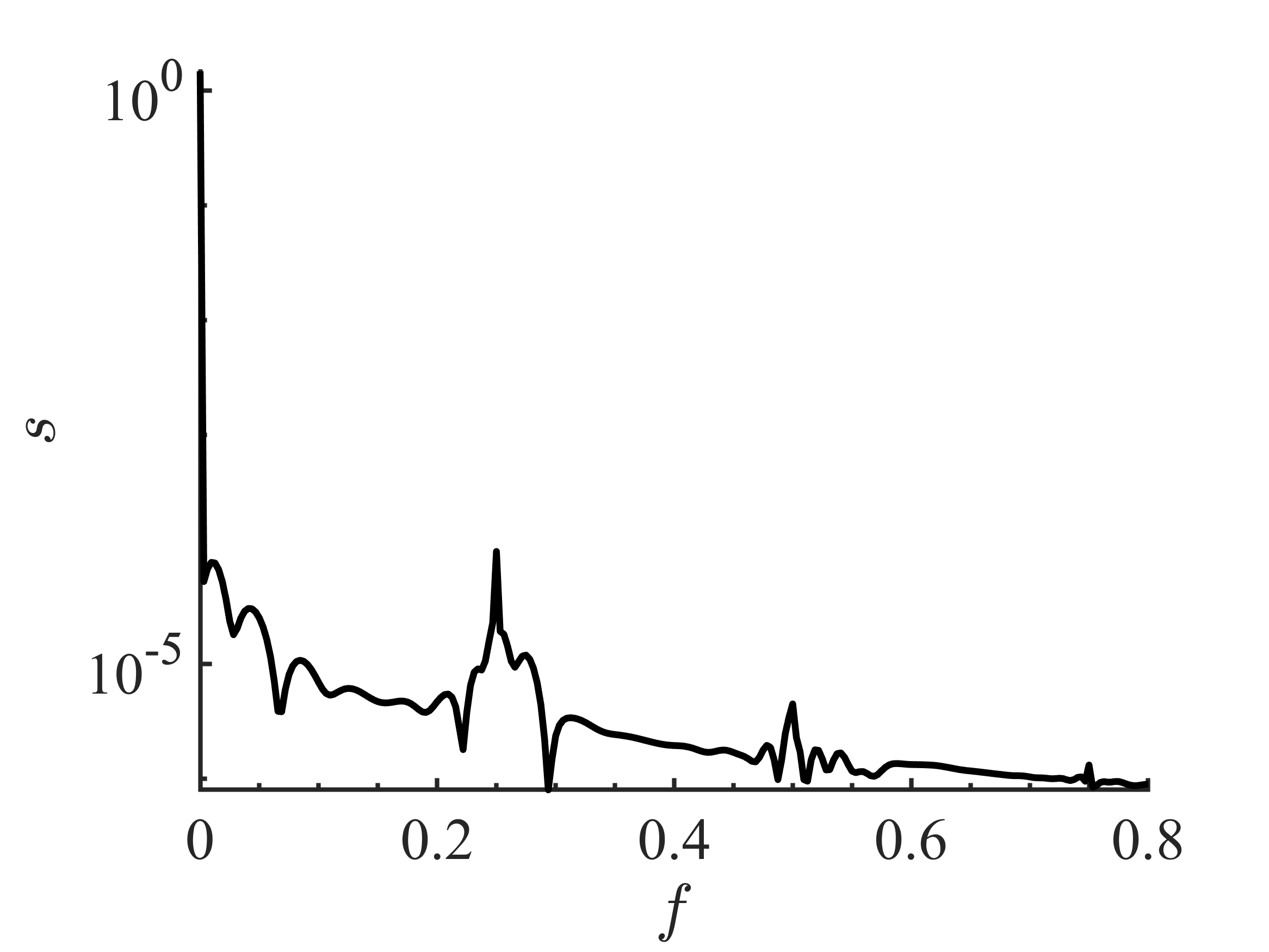}\label{fig:frequency-Ra-2100_2-subfig}} \hfill
\subfloat[Spectrum of periodic behaviour.]{\includegraphics[width=0.5\textwidth]{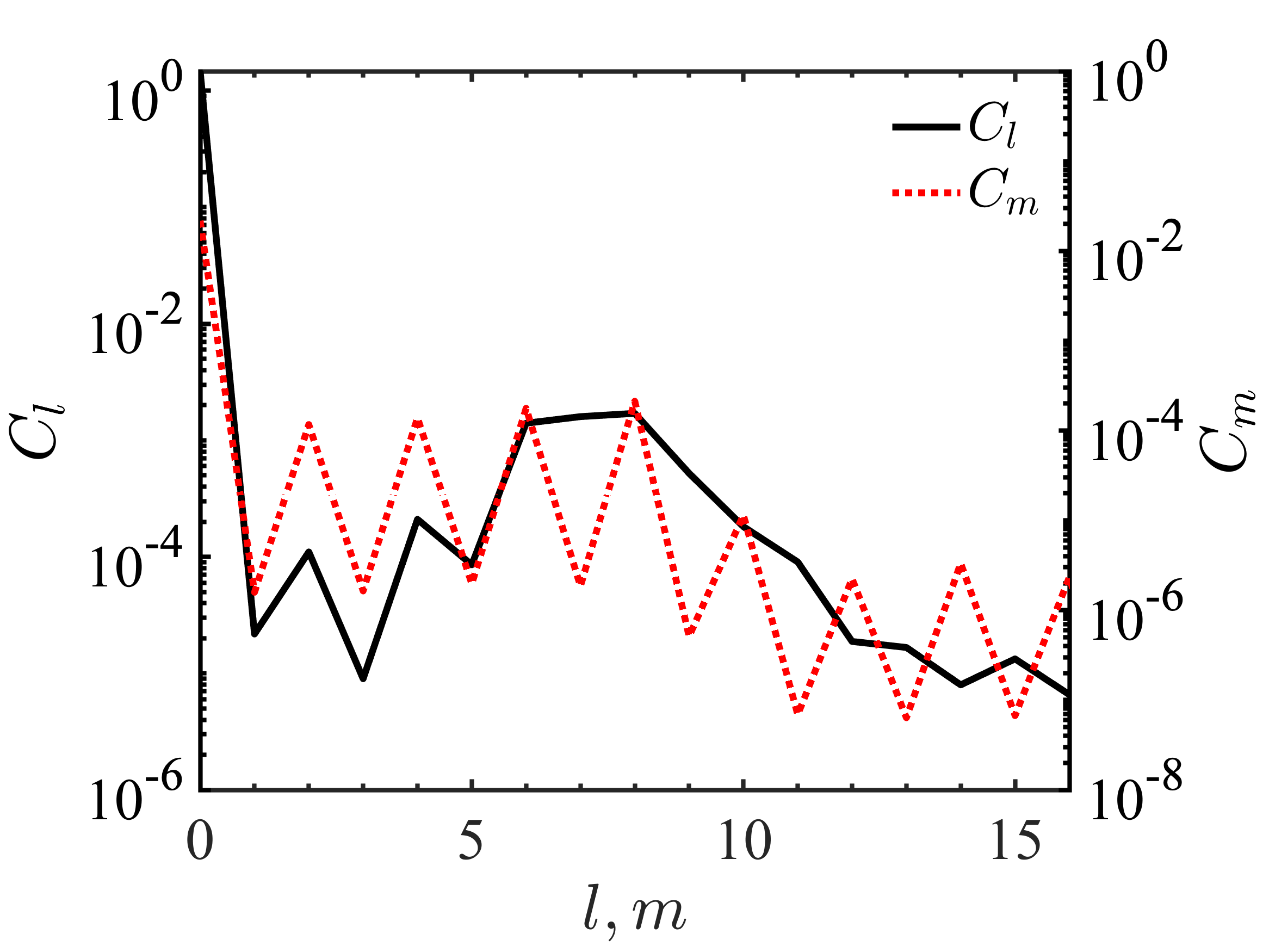}\label{fig:spectrum-Ra-2100_2-subfig}}
\caption{Analysis of quadratic gravity at $Ra_2=2100$. Legend for temperature and spectrum profiles as in figure \ref{fig:spectrum_convection}.
}
\label{fig:Ra_2_full}
\end{center}
\end{figure}

When $Ra$ is increased past $Ra_2$ a new behaviour appears. At first, there is a transitional phase where $Nu(t)$ does not follow a predictable pattern; these oscillations are then replaced by a periodic dynamics that remains stable at any greater $t$. In figure \ref{fig:Ra_2_full} the behavior for $Nu(t)$ at $Ra=2100$ for quadratic gravity is shown. In the instantaneous snapshot of the temperature profile (figure \ref{fig:temperature-Ra-2100_2-subfig}) we can notice how plumes on the x-y plane are now significant. This phenomenon is also evident by looking at the spectrum profile (figure \ref{fig:spectrum-Ra-2100_2-subfig}): while $C_l$ maintains approximately the same magnitude as lower $Ra$ simulations (with a shift in main-degree to the range $6-8$), $C_m$ is now significantly greater than $0$ even for $m\neq 0$ and presents the same alternating structure that characterised $C_l$ before. Indeed, in this region the flow is not stationary; thus, the degeneracy of the eigenvalues is lost. In figure \ref{fig:frequency-Ra-2100_2-subfig} the Fourier transformation of the oscillatory evolution is shown (for further details, see Appendix \ref{app_spectra}): two main frequencies can be highlighted, $f=0.25$ (main peak) and $f=0.5$. 

\begin{figure}
\begin{center}
\subfloat[$Nu$ vs $t$ at $Ra=5000$. Inset: zoom on the periodic behaviour.]{\includegraphics[width=0.5\textwidth]{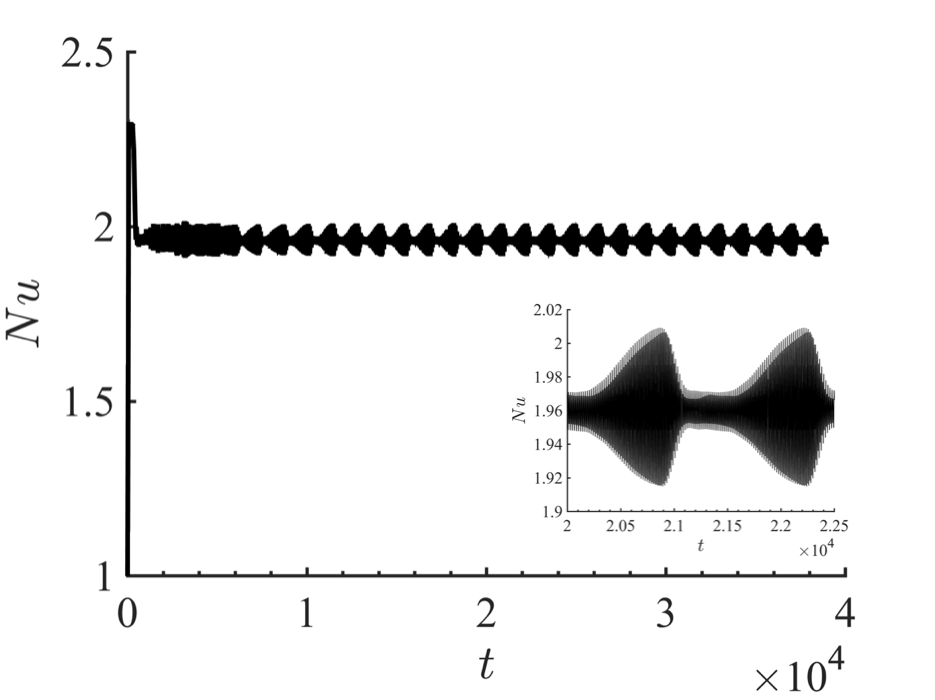}} \hfill
\subfloat[Intensity $s$ vs frequency $f$ for $Ra=5000$.]{\includegraphics[width=0.5\textwidth]{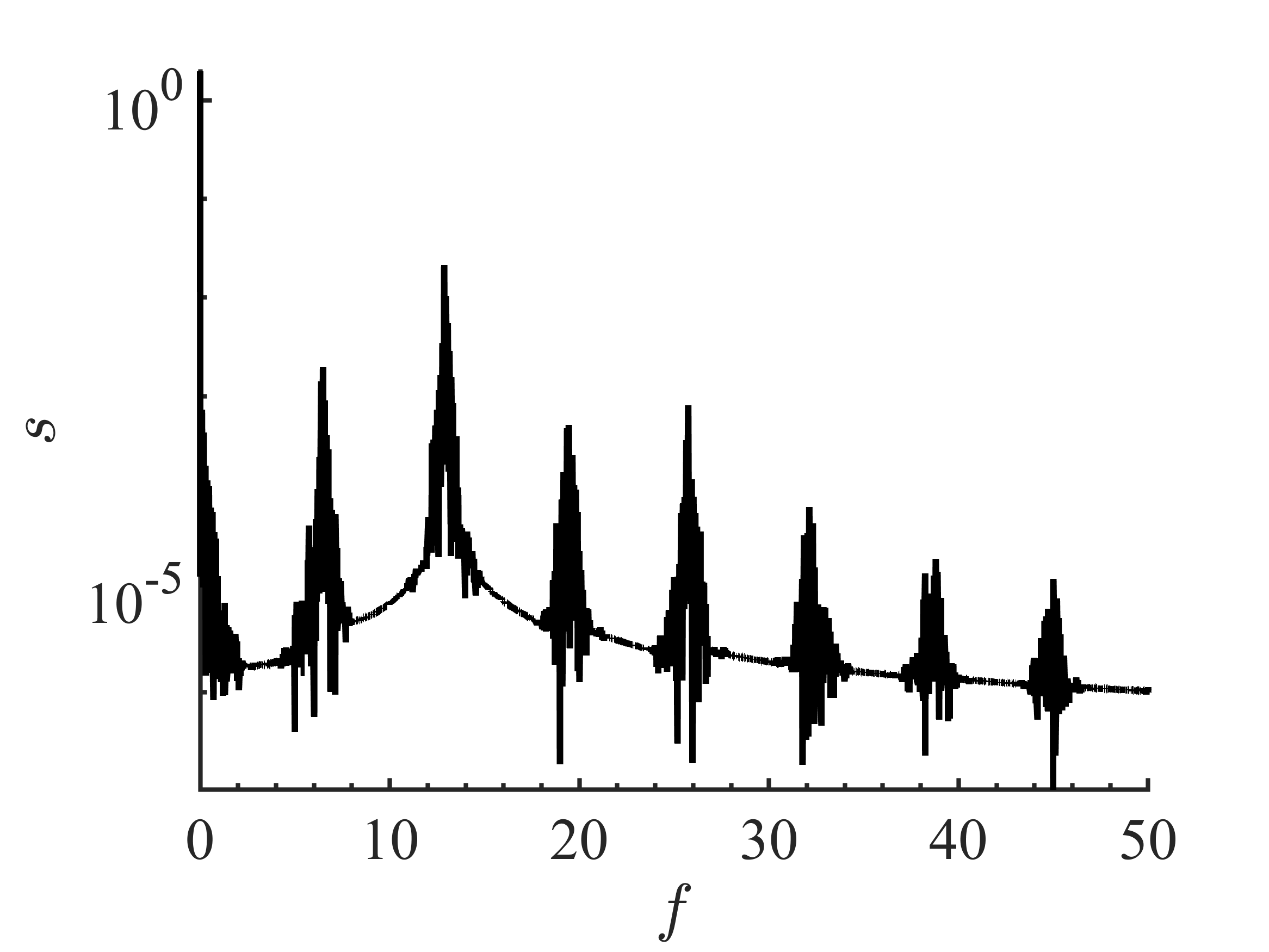}} \\[-2ex]
\subfloat[$Nu$ vs $t$ at $Ra=10000$.]{\includegraphics[width=0.5\textwidth]{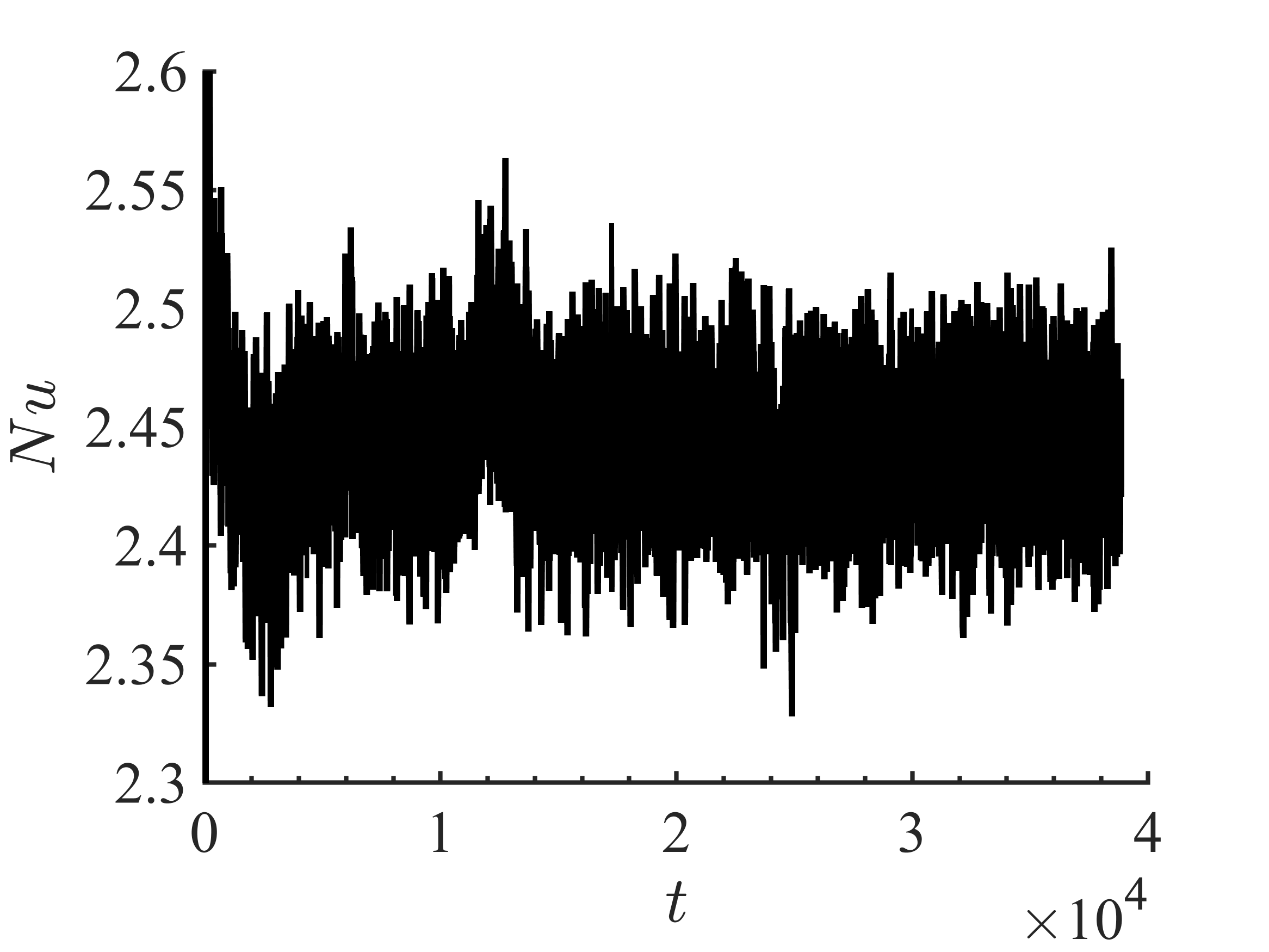}} \hfill
\subfloat[Intensity $s$ vs frequency $f$ for $Ra=10000$.]{\includegraphics[width=0.5\textwidth]{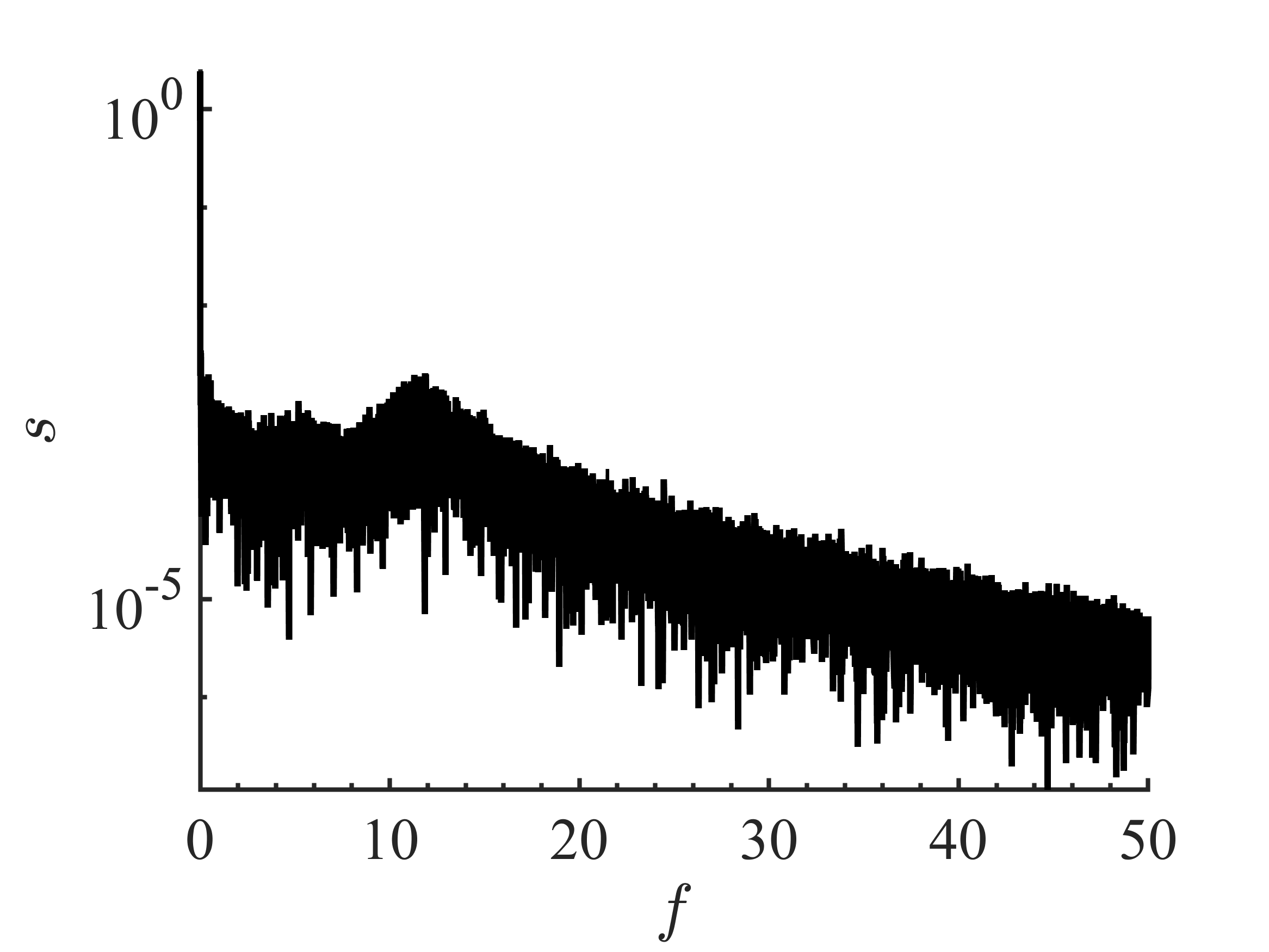}} \\[-2ex]
\subfloat[Snapshot of \\temperature profile for \mbox{$Ra=5000$}.]{\includegraphics[width=0.22\textwidth]{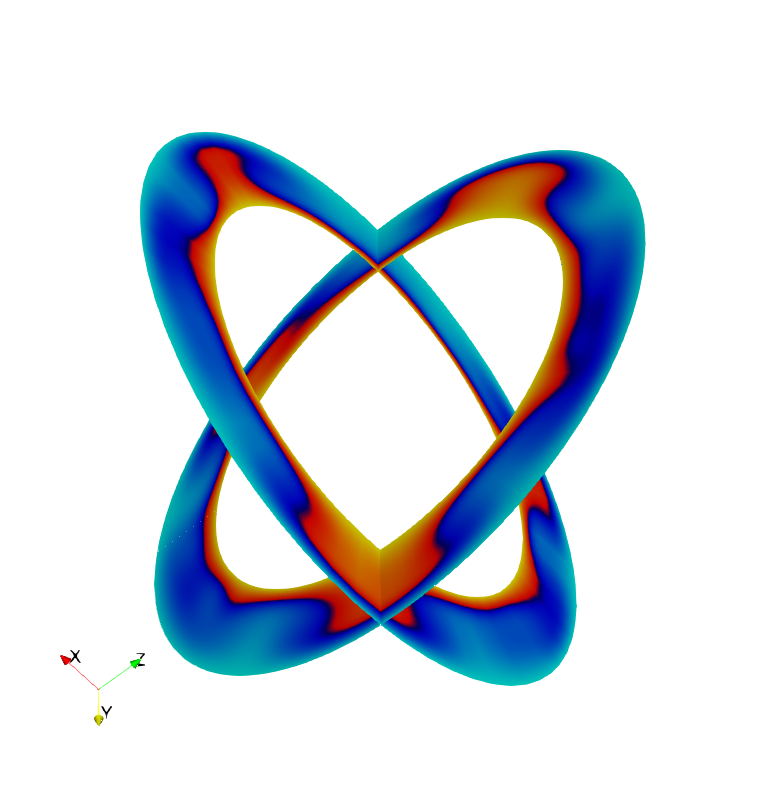}} \hfill
\subfloat[Harmonic spectrum of  {$Ra=5000$}.]{\includegraphics[width=0.25\textwidth]{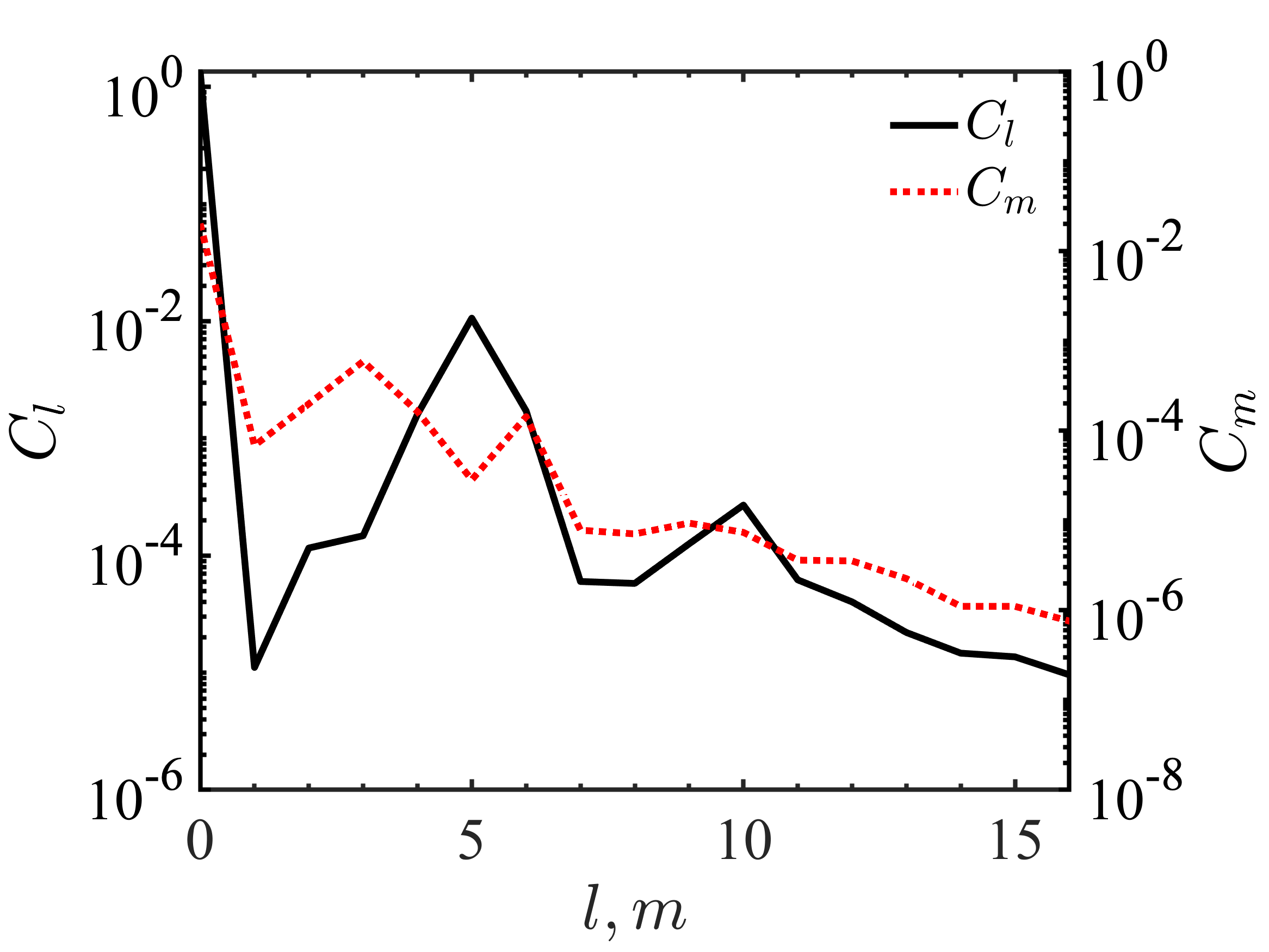}} \hfill
\subfloat[Snapshot of \\temperature profile for \mbox{$Ra=10000$}.]{\includegraphics[width=0.22\textwidth]{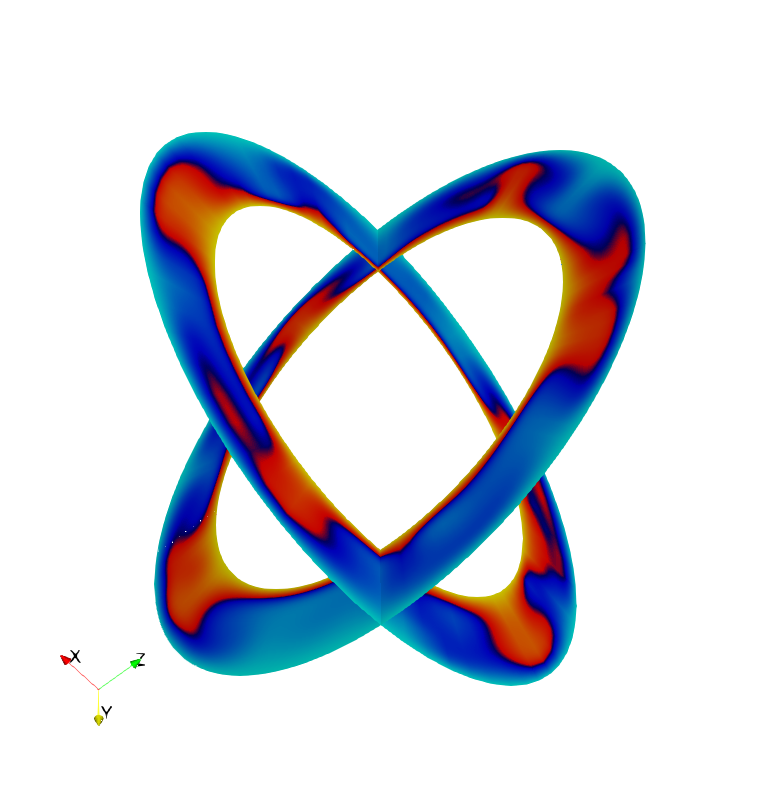}} \hfill
\subfloat[Harmonic spectrum of  $Ra=10000$.]{\includegraphics[width=0.25\textwidth]{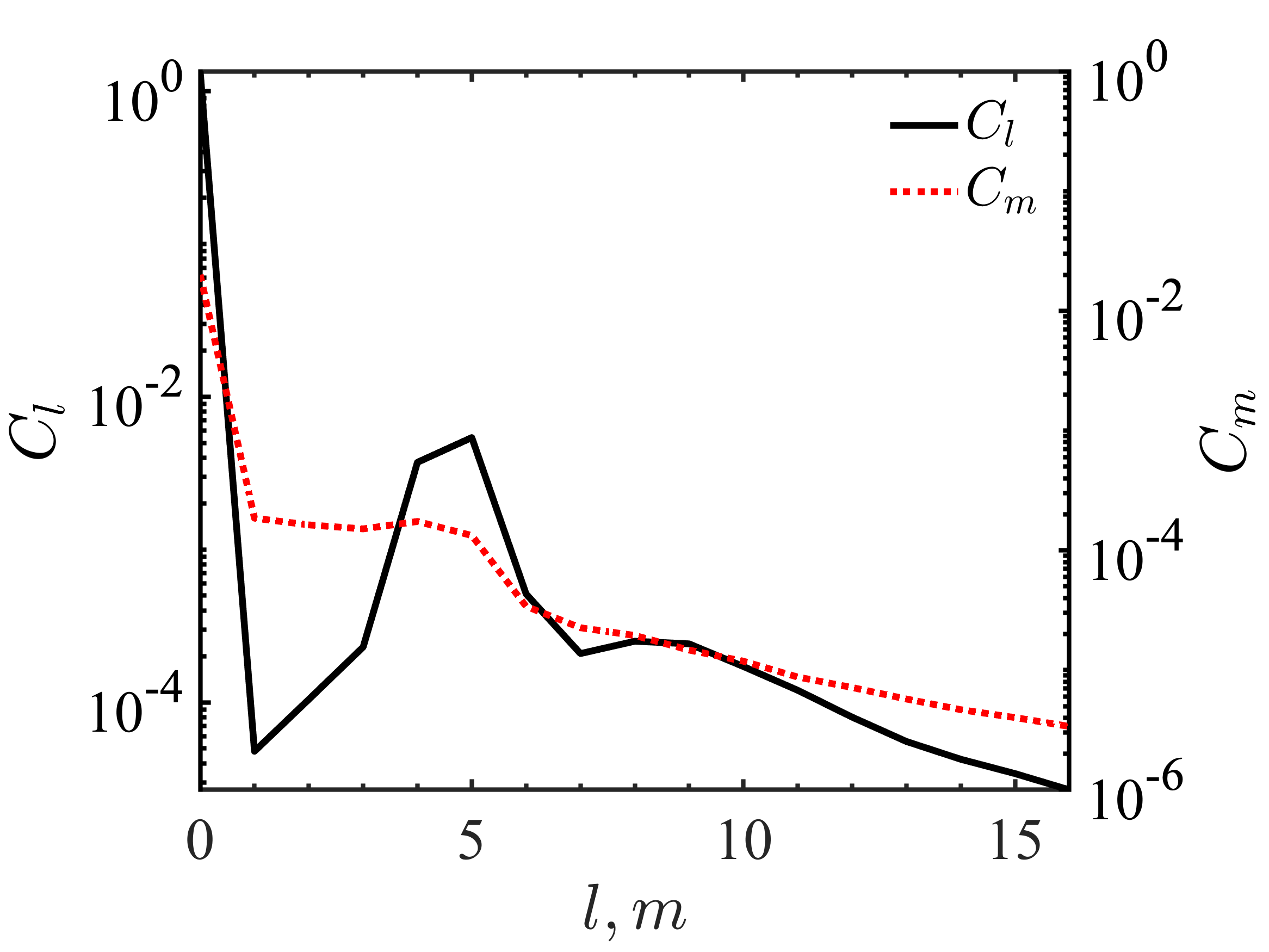}}
\caption{Analysis of quadratic gravity at $Ra=5000$ (a,b,e,f) and $Ra=10000$ (c,d,g,h).
Legend for temperature and spectrum profiles as in figure \ref{fig:spectrum_convection}.
}
\label{fig:Ra5000}
\end{center}
\end{figure}

Further increases of $Ra$ reduce the transitional period and introduce new frequencies in the spectrum; In figure \ref{fig:Ra5000}, the case for $Ra=5000$ for quadratic gravity shows that the harmonic spectrum has magnitude closer to $\mathcal{O}(1)$ for both $C_l$ and $C_m$, with a main-degree at $5$, and the frequency spectrum shows several peaks. At higher values of $Ra$, periodicity disappears and a chaotic behaviour is obtained: here the frequency spectrum is almost continue. Surprisingly, the main-degree remains stable around $l=5$ instead of following the previous behaviour of shifting towards lower numbers for higher values of $Ra$. $C_m$ remains highly excited for a large range of $m$.

Region II is then defined as the region for which the system presents a time--dependent behaviour at any time $t$. This behaviour may be periodic when $Ra$ is very close to $Ra_2$, or almost chaotic when $Ra$ is much higher.

\subsection{Hysteresis}

\label{sec:hysteresis_rbc}

\begin{figure}
\begin{center}

\subfloat[$Nu$ vs $t$ for increasing and decreasing evolutions. The difference between the two approaches is $\approx 2\%$.]{\includegraphics[width=0.47\textwidth]{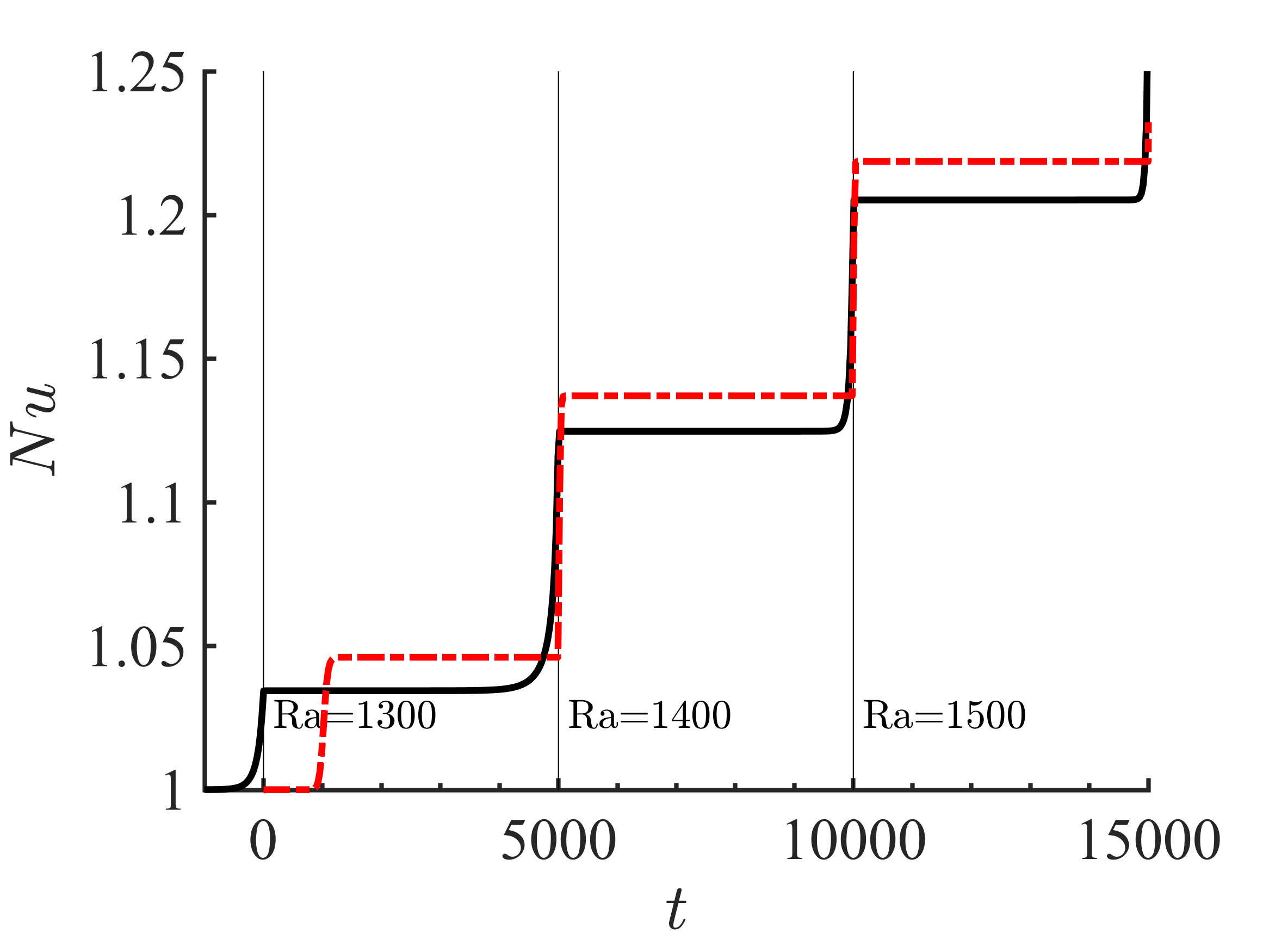}\label{fig:increasingVSdecreasing}} \hspace{3mm}\hfill \rulesep \hfill 
\subfloat[$Ra$ vs $t$ around $Ra_2$. In the inset, zoom on the periodic behaviour of black case.]{\includegraphics[width=0.47\textwidth]{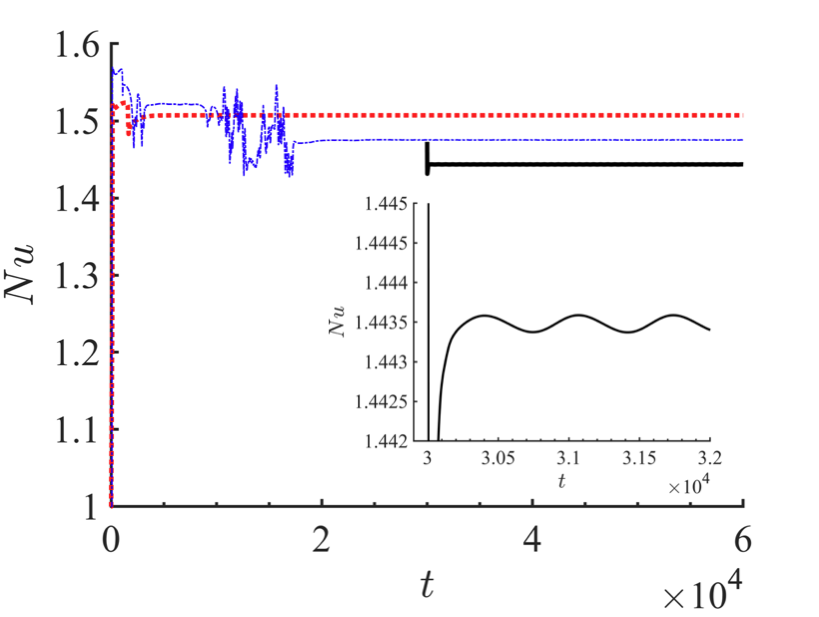}\label{fig:Ra2000NuVSt}}\\[-5ex]
\subfloat[$Nu$ vs $t$ for two different decreasing cases. In the inset, zoom: the difference between the two cases is less than $0.2\%$ here.]{\includegraphics[width=0.47\textwidth]{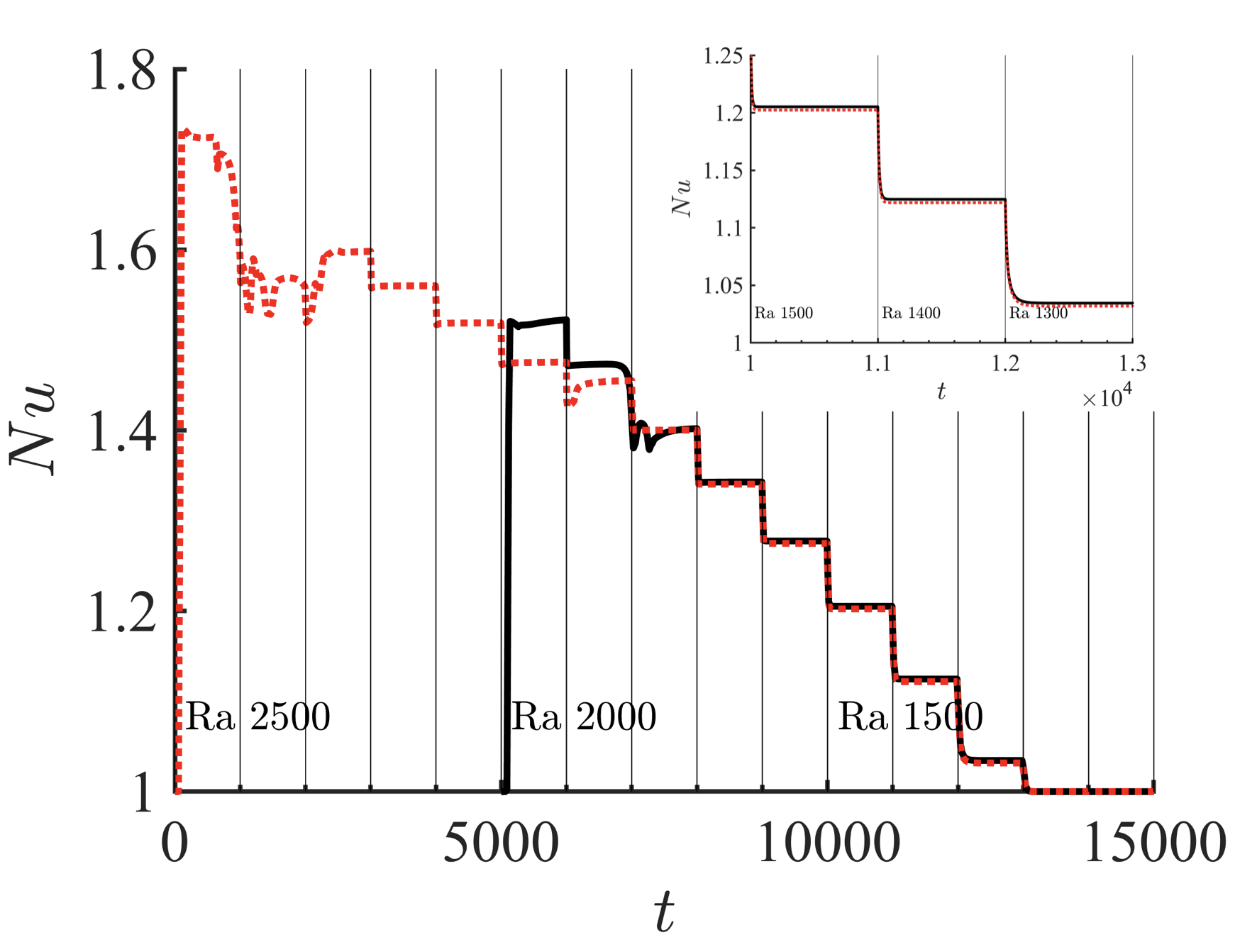}\label{fig:decreasing1VS2}}
\hspace{3mm}\hfill \rulesep \hfill 
\subfloat[Snapshot of temperature profile for inset of (b). Behaviour closer to $Ra_2$ (fig \ref{fig:temperature-Ra-2100_2-subfig}) than $Ra_1$ (fig \ref{fig:temperature_Ra-1700_8-subfig}). ]{\includegraphics[width=0.42\textwidth]{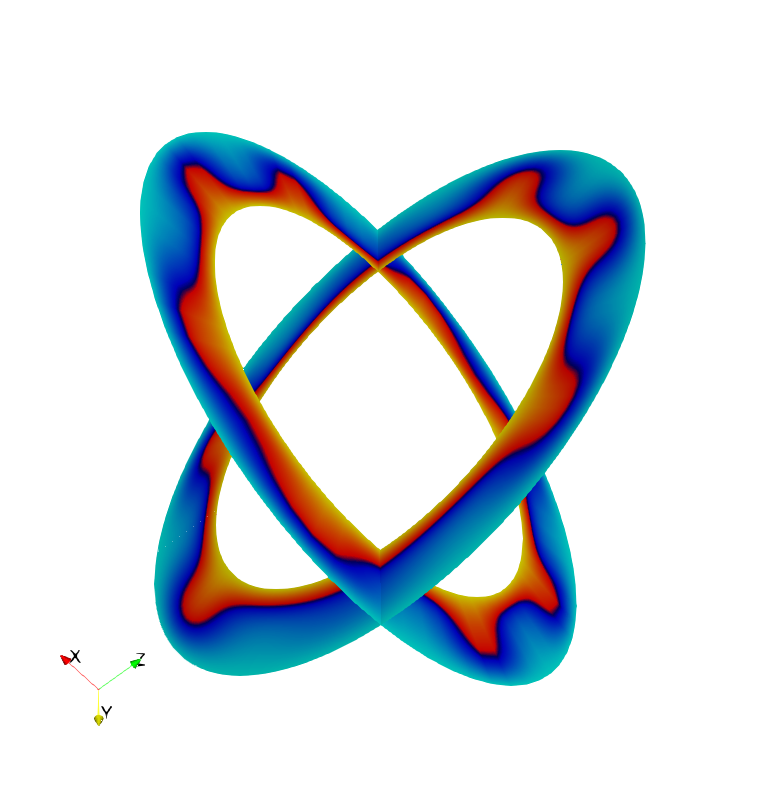}\label{fig:Ra2000temperature}}\\[-2ex]
\subfloat[Harmonic spectrum for red line of (c), at $Ra=1300$.]{\includegraphics[width=0.47\textwidth]{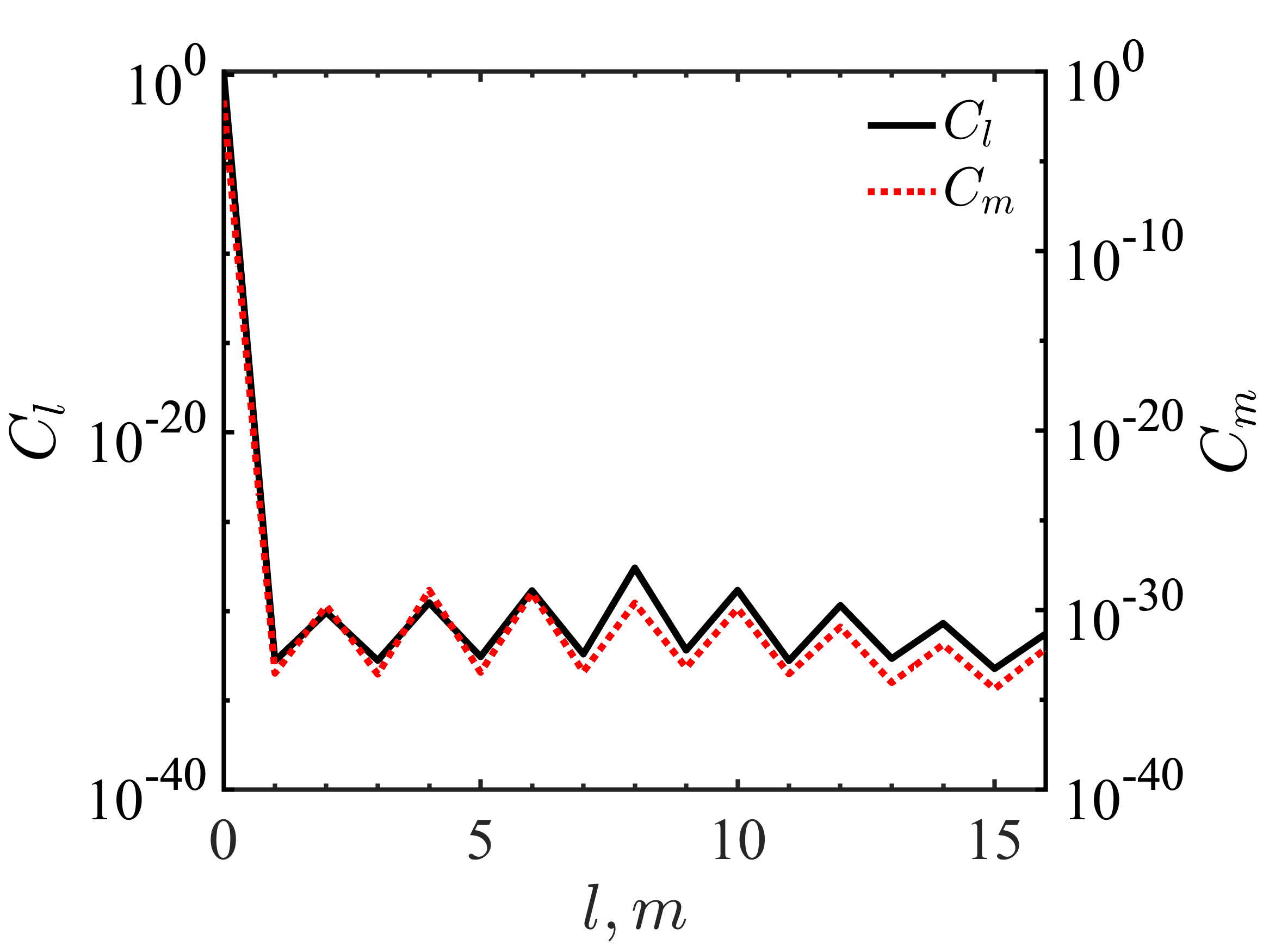}\label{fig:decreasing2spectrum}}
\hspace{3mm}\hfill \rulesep \hfill 
\subfloat[Harmonic spectrum for inset of (b). Behaviour closer to $Ra_2$ (fig \ref{fig:spectrum-Ra-2100_2-subfig}) than $Ra_1$ (fig \ref{fig:spectrum_Ra-1700_8-subfig}).]{\includegraphics[width=0.47\textwidth]{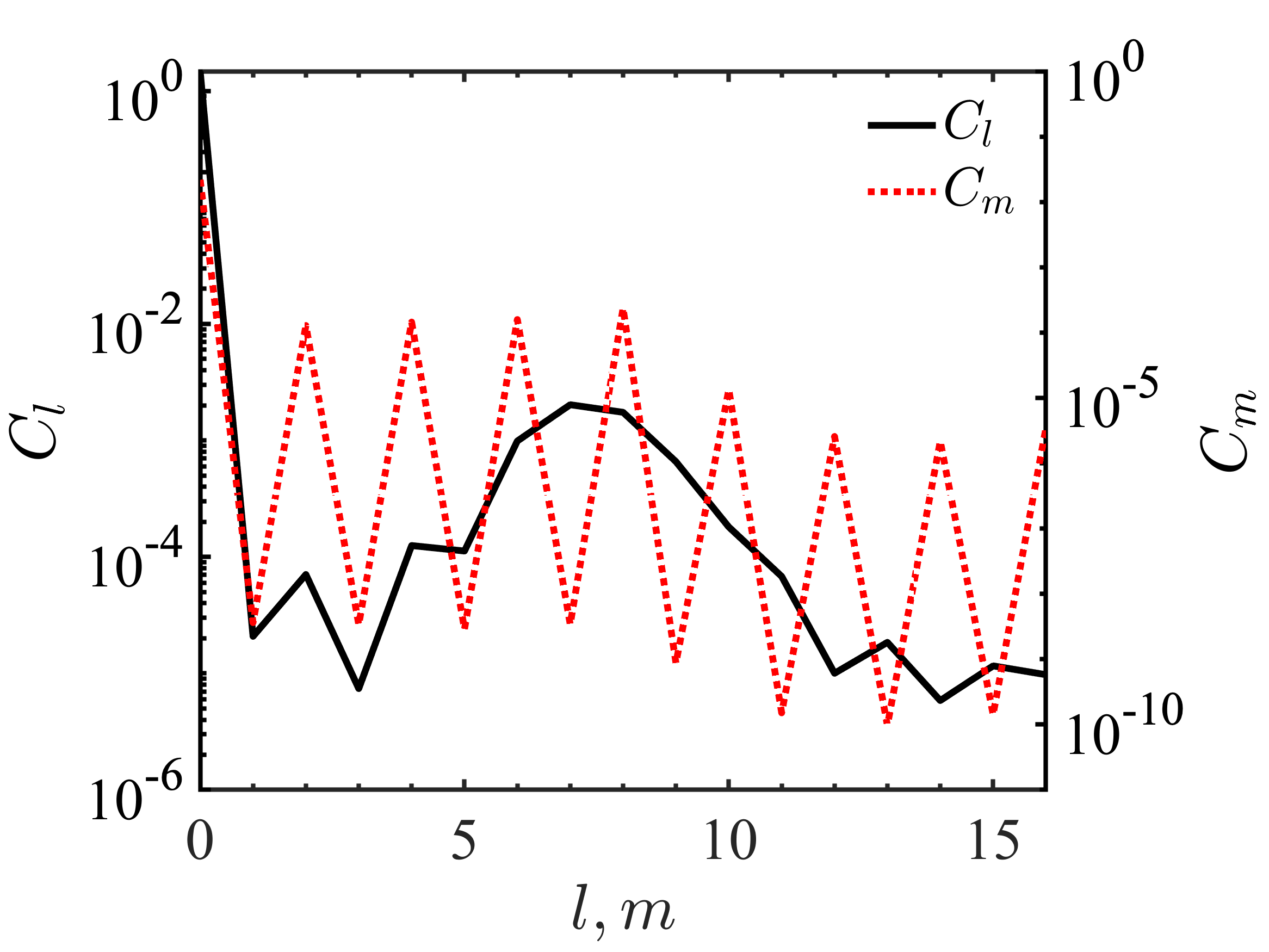}\label{fig:Ra2000spectrum}}
\caption{Left side: initial conditions comparison (vertical black lines in (a,c) indicate changes in $Ra$); right side: behaviour around $Ra_2$. (a): %
{\color{black} \solid }decreasing from $Ra>Ra_1$ (time scale reversed; %
{\color{red} \dotted }increasing from $Ra_c$.  %
(b): %
{\color{black} \solid }\textit{oscillating state}; %
{\color{red} \dotted }\textit{rest state}; %
{\color{blue} \chndash}\textit{region II state}. %
(c): %
{\color{black} \solid }decreasing from $Ra<Ra_2$; %
{\color{red} \dotted }decreasing from $Ra>Ra_2$. %
Legend for temperature and spectrum profiles as in figure \ref{fig:spectrum_convection}.
}
\label{fig:hysteresis_ra}
\end{center}
\end{figure}

As anticipated in previous sections, the behaviour of the system varies accordingly to the starting conditions. This is in line with the analysis of \citet{mannix2019weakly}, which predicted the stable solution to have a dependency on starting conditions, as well as the impact of $l\pm2$ and $l\pm1$ modes. From current analysis, we identified two main time-independent states, $\mathcal{S}_9$ and $\mathcal{S}_8$. The stability analyses of \citet{araki1994thermal} and \citet{Avila2013} for $R_i=2.5$ suggest that, due to a degeneracy in the solutions, the onset of convection can occur for both $l=8$ and $l=9$, which corroborates our results.  
In figure \ref{fig:increasingVSdecreasing}, we compared the convection efficiency of  heat transfer for two different simulations: an increasing case starting from the onset of convection, and a decreasing case starting in region I and with enough time to reach state $\mathcal{S}_8$. Results show that the second case has a worse efficiency (i.e. a smaller value of $Nu$): an explanation for this effect can be found in the temperature profile, which for the $\mathcal{S}_8$ state has less protrusions between the two shells and thus a reduced heat transfer. At the present moment, however, it is unclear why the system prefers a configuration with a less efficient heat transfer. A possible interpretation of these results come from the comparison with \citet{araki1994thermal}, which show that for this $R_i$, lower modes $l$ correspond to higher values of $Ra$; thus, it is possible that once a sufficiently high $Ra$ is reached, the system sits on the higher curve of $l=8$ for its critical value, while for lower explored values of $Ra$ the only available curve is for $l=9$.

It is worth noticing that using increasing or decreasing simulations is just a matter of convenience: as long as the initial conditions of the system are in state $\mathcal{S}_9$ and the system does not evolve in $Ra>Ra_1$, the dynamic is fully equivalent to the increasing case with initial conditions at $Ra_c$.

A similar result is obtained when looking at the time dependent behaviour of $Ra\ge Ra_2$. Figure \ref{fig:Ra2000NuVSt} shows the behaviour of the simulations under three different conditions: an evolution in region II at $Ra_2$ (we name it \textit{region II state}), and two evolutions at $\Bar{Ra}=2000<Ra_2$ differentiated by their initial conditions. Of these two, the first is done at initial conditions $\Bar{Ra}$ (\textit{rest state}), the other starts from the periodic oscillations of the region II simulation, i.e. initial conditions $Ra_2$ (\textit{oscillating state}). This latter simulation keeps in its evolution the periodic pattern presents in \textit{region II state} and typical of higher values of $Ra$. Its spectrum is qualitatively identical to the spectrum of the \textit{region II state}, as shown by comparing figures \ref{fig:Ra2000spectrum} and \ref{fig:spectrum-Ra-2100_2-subfig}, and the same holds true for the temperature profile, showed in figures \ref{fig:Ra2000temperature} and \ref{fig:temperature-Ra-2100_2-subfig}. On the other side, the spectrum and temperature profile of the \textit{rest state} (not shown here) is qualitatively equivalent to any other fluid evolved completely in region I. 

The periodic oscillations remain a feature of the system even when $Ra$ is further decreased, but their effect on $Nu$ becomes progressively smaller and it is negligible when $Ra\lesssim Ra_1$, as shown by figure \ref{fig:decreasing1VS2}: the difference between two evolutions with decreasing $Ra$ is large at the beginning, but becomes almost zero when $Ra$ approaches $Ra_c$. Even if the effect on heat transfer is negligible, the system keeps partial memory of its initial conditions in the spectrum: in a simulation with oscillating initial conditions the spectrum has a non-zero oscillating value of $C_m$ even for very low $Ra$, being $5$ orders of magnitude bigger in figure \ref{fig:decreasing2spectrum} than the comparable simulation at $Ra\approx Ra_c$ of figure \ref{fig:profile_very_low_Ra}. 

While in general we expect to see $Nu(Ra=x)>Nu(Ra=y)$ when $x>y$, we can notice that in our simulations this order is not respected when comparing $Nu(Ra=Ra_2)$ and $Nu(Ra=\Bar{Ra}_{rest})$ (figure \ref{fig:Ra2000NuVSt}). As previously noted, the rise of a periodic oscillation and the change in the spectrum are correlated to a worse heat transfer, this justifies the inversion in $Nu$. Indeed, when comparing the two oscillating solutions, we restored the expected order: $Nu(Ra=Ra_2) > Nu(Ra=\Bar{Ra}_{oscillating})$.

The strong dependence on the initial conditions seem to be an intrinsic property of the system that does not depend on the gravity profile chosen: preliminary tests have been run for all the other profiles and, as already shown by table \ref{tab:ra_1and2}, the existence of those regions is not affected by the chosen gravity.

\subsection{Higher Pr Number}
\label{sec:water_rbc}

As anticipated in section \ref{sec:numsim_rbc}, the same configurations previously analysed have been studied also for water, characterised by $Pr_{water}=7.1=10Pr_{air}$. Looking at equation \eqref{eq:NavierStokes}, we notice how the main direct effect of a variation in Prandtl number is a larger coefficient in front of $\laplacian\vb{u}$, i.e. the viscous effects are stronger for water compared to air. Thus, for inertial terms to overcome viscosity, an higher $Ra$ is needed. Since $Pr_{water}/Pr_{air}=10$, we expect $Ra_{water}$ to be $10$ times bigger to obtain the same time-dependent behaviours, while the critical value for the onset of convection will be unaffected by the change of $Pr$.

The system has been tested for different grids, and the same grid used for simulations at lower $Pr$ has been found to be accurate.
The \textit{increasing / decreasing} approach has been tried for this value of $Pr$, but it has been found rather ineffective due to the much larger time needed to reach a stable state, especially for the \textit{increasing} case, thus the study has been done by running several simulations, each at a (different) fixed $Ra$. In most of the simulations the fluid is initialised from rest, but to analyse hysteresis and for longer simulations sometimes an old state has been used as starting condition for the new run.

Using $\vb{g}^q$ as gravity profile the onset of convection, computed as before by the analysis of the Nusselt number, happens at the same value of $Ra_c \approx 1250$, equivalent to an effective value of $Ra_c^e=1750$. This result, in line with the predictions, shows that the onset of convection has no direct dependence on the value of $Pr$. When $Ra$ is much higher than $Ra_c$, we observe again the time dependent behaviour of region II appearing at $Ra_{2, water}\approx 21000$, that respects our prediction of $Ra_{2,water}=10Ra_{2, air}$.

In the region I between the onset of convection and the time dependent behaviour, a series of different regimes appear. Looking at the spectrum, we find the same configuration found for air, with a state $\mathcal{S}_9$ having at spectrum peaked at $C_l=9$. This state is stable for a much larger range of $Ra$ compared to air but eventually, as $Ra \approx 4000=Ra_1$, the system jumps to a new equilibrium state $\mathcal{S}_8$, with a spectrum peaked at $C_l=8$. In the same fashion as air, decreasing $Ra$ from a system in $\mathcal{S}_8$ does not bring it back to $\mathcal{S}_9$: hysteresis is present also for water. Compared to air simulations, however, the system has a larger space to explore in $Ra$ before reaching region II at $Ra_2$. This leads to the appearance of new states in the meta-stable region I, we identify them as $\mathcal{S}_7$, $\mathcal{S}_6$ and $\mathcal{S}_5$ for systems with main-degree $7$,$6$ and $5$ respectively. A summary of the first measured $Ra$ for each state is given in table \ref{tab:water_table}. As happened for air, when the main peak is an odd/even number then the all the odd/even degrees are peaked as well. As noted in the previous section for air, we believe that the curves of $Ra(l)$ shown in \citet{araki1994thermal} can offer an explanation about this behaviour.

Increasing $Ra$ above $Ra_2$ brings back the same periodic oscillation observed for region II of air, with the harmonic spectrum keeping its peak at $l\approx 5$ and the rise of more and more characteristic frequencies. Surprisingly, the state $\mathcal{S}_5$ appears to be the most stable both for water (which reaches it after exploring all the states from $9$ to $5$ in region I) and for air (which reaches it only in region II). Eventually, for $Ra$ high enough, the frequency spectrum becomes continuum and the system moves toward a turbulent state. 

\begin{table}
    \centering
\begin{tabular}{c||c|c|c|c|c|c}
    \hline
&$Ra_{\mathcal{S}_9}$ & $Ra_{\mathcal{S}_8}$ &$Ra_{\mathcal{S}_7}$ & $Ra_{\mathcal{S}_6}$ & $Ra_{\mathcal{S}_5}$  & $Ra_2$ \\
\hline
$Ra^q$&$1250$  & $4000$      & $6000$        & $11000$       & $16000$       &  $21000$\\
$Ra^e$&$1740$  & $5550$     &   $8350$      &   $15300$     &   $22300$     &   $29000$
\end{tabular}
\caption{Values of $Ra$ for the first occurrence of different states when $Pr=7.1$ for quadratic gravity on the first line, and effective value on the second line. $Ra_{\mathcal{S}_9}$ is equivalent to $Ra_c$ and $Ra_{\mathcal{S}_8}$ is equivalent to $Ra_1$. Confidence interval at $\pm5\%$ for $Ra_c$ and $Ra_2$, $\pm10\%$ for the others.}
\label{tab:water_table}
\end{table}

\section{Conclusion}
\label{sec:conclusion_rbc}

In this paper, a characterization of Rayleigh-Bénard convection for fluids between spherical shells has been carried out using a three-dimensional second order finite difference scheme in spherical coordinates. By setting fixed temperature at the shells and fixed radius ratio, we can explore different configurations by varying Prandtl number, Rayleigh number, and the radial gravity profile. 

The results for the onset of convection are compared with linear stability analysis, which predicts a zero-order critical Rayleigh number $Ra^{(0)}\approx1708$. The study up to first order correction for different gravity profiles yields various results: while for constant and linear gravity profiles the difference between data and theory is very small, the quadratic and mantle-like profiles differ for up to $10\%$ from the expected value. However, perfect agreement is restored when an averaged value $Ra^e\approx1730$ is computed: thanks to it the first order correction to $Ra^{(0)}$ vanishes and the discrepancy between the measured values and the linear analysis is kept below $1\%$. This result holds true for both air ($Pr_{air}=0.71$) and water ($Pr_{water}=7.1$), which is an expected behaviour, given that no dependency on $Pr$ is present. 


The effective Rayleigh number can also be used to identify all the subsequent states the system explores. Our criterion for the characterization of different states is given by the analysis of the spherical harmonics of the system. At the onset of convection, the system has a harmonic spectrum peaked at degree $l=9$, so we identify this state as $\mathcal{S}_9$. Increasing $Ra$ leads to the rise of new situations. For air, we first identify a region I (for $Ra^e\ge Ra^e_1=2350$) where the state $\mathcal{S}_9$ becomes unstable and the system, given enough time, moves to a new configuration $\mathcal{S}_8$ where the main-degree of the harmonic spectrum is $l=8$ ($C_m$ remains unexcited in this region). Region I for water starts at a higher Rayleigh number ($Ra^e_1\approx 5550$) but the system remains in this region for a larger interval of $Ra$; therefore, while both fluids start from state $\mathcal{S}_9$ and reach state $\mathcal{S}_8$, increasing $Ra$ for water leads to the emergence of new states $\mathcal{S}_7, \mathcal{S}_6$ and $\mathcal{S}_5$ where the main-degree is, respectively, $7,6$ and $5$.

When $Ra$ is increased beyond a threshold $Ra_2$, the system enters region II, where a time--dependent behaviour is observed. Being this behaviour heavily influenced by $Pr$, we expect $Ra_{2, water}$ to be about $10$ times bigger than $Ra_{2,air}$; indeed we have $Ra_{2,air}^e=2900$ and $Ra_{2, water}^e=29000$. In this region, the harmonic spectrum has main-degree $l=5$ at any $Ra$ for both water and air, while the values of $C_m$ become progressively larger. In this region the analysis of the frequency spectrum can give interesting information. For $Ra$ close to $Ra_2$, only few frequencies are excited and a clearly periodic behaviour can be observed in the flow dynamic. Increasing $Ra$ leads to new peaks in the frequency spectrum and eventually a continuum spectrum is attained at very high values of $Ra$, while the periodicity in the dynamic disappears.

We observed that, for both water and air, lower degree states are stable, and decreasing $Ra$ to previously explored values does not bring back the higher--degree configurations. Moreover, we discovered that lower $l$ is associated with a reduced heat transfer and thus a lower value of $Nu$. We also noted that, if the system explored region II during its evolution, time dependent features remain part of the dynamic even when $Ra$ is lowered, and $C_m$ remains excited even for values of $Ra$ very close to the onset of convection.
Based on these observations, we can claim that the system presents hysteresis and its heat transfer is heavily dependent on its starting conditions.

\section*{Acknowledgements}
We acknowledge the support of the GNFM group of INDAM and the national e-infrastructure of SURFsara, a subsidiary of SURF cooperation.

\appendix
\section{Appendix: Harmonic analysis and spectrum}\label{app:spectrum}\label{app_spectra}
This appendix describes the spectral analysis procedure used in this study. 
\subsection*{Fourier decomposition}
Fourier transforms are a fundamental tool in mathematics and physics. A Fourier transform of a function of time is a (in general complex valued) function of frequency, and its magnitude or intensity gives information to the most \textit{common} frequencies of the original function. Let $f$ a periodic square--integrable function $\hat s  \in L^2(T)$, with T being the unitary circumference. Then its Fourier series is
\begin{equation}
    \sum_{n=-\infty}^{n=\infty}s(n)e^{int}
\end{equation}
where $i$ is the imaginary unit and the Fourier coefficients $s(n)$ are defined as
\begin{equation}
    s(n)=\frac{1}{2\pi}\int_{-\pi}^{\pi}\hat s (t)e^{-int}\dd t = \frac{1}{2\pi}\int_{-T/2}^{T/2}\hat s (t)e^{-i2\pi n t /T}\dd t.
\end{equation}
The coefficients can be seen as discretized samplings of the Fourier transform at intervals $1/T$, so we can define the Fourier transform $s(f)$ as 
\begin{equation}
    s(f)=\frac{1}{2\pi}\int_{\mathbb{R}}\hat s(t)e^{-ift}\dd t.
\end{equation}

\subsection*{Spherical harmonic decomposition}
Given a square--integrable function $f:S^2\to\mathbb{C}$ on the unit sphere $S^2$
its spherical harmonic decomposition can be written as 
\begin{equation}
    f(\phi, \theta)=\sum_{l=0}^\infty\sum_{m=-l}^l C_l^m Y_l^m(\phi, \theta)
    \label{eq:appendix_spherical-harmonic-decomposition}
\end{equation}
where $Y_l^m(\phi, \theta)$ is the spherical harmonic of degree $l$ and order $m$ (which represent the wavenumber along a meridian and the equatorial plane), and $C_l^m$ its coefficient, and the expansion holds in the sense of convergence in $L^2$ of the sphere.
$Y_l^m(\phi, \theta)$ can be defined in terms of associated Legendre polynomials $P_l^m$ by 
\begin{equation}
    Y_l^m(\phi, \theta) = \sqrt{\frac{(2l+1)(l-m)!}{4\pi(l+m)!}} P^m_l(cos\phi)e^{im\theta}, 
    \label{eq:orthonormalised_harmonic}
\end{equation}
where $P_l^m(x)$ satisfies the general Legendre equation
\begin{equation}
    \dv{x}\qty[(1-x^2)\dv{x}P^m_l(x)] + \qty[l(l+1)-\frac{m^2}{1-x^2}]P_l^m(x)=0.
\end{equation}

For orthonormalised harmonics, as the one defined in equation \eqref{eq:orthonormalised_harmonic}, the coefficients can be computed by
\begin{equation}
C_l^m = \int_\Omega f(\phi, \theta) Y_l^{m*}(\phi,\theta)\dd \Omega    
\label{eq:appendix_Clm1}
\end{equation}
where $\Omega$ is the solid angle. 

This tool has been used in this study to perform a spectral analysis of the averaged square temperature $T^2(\phi, \theta)$ to better characterize the fluid behaviour for the various simulations. Coefficients of equation \eqref{eq:appendix_Clm1} are obtained by using the SPHEREPACK library \citep{adams1999spherepack}.
In the analysis, the averaged value of coefficients has been used, i.e. $C_l=c_0 \langle \sum_{m=-l}^l C_l^m \rangle$ (with $c_0$ normalization factor) and
\begin{equation}
    C_{m}= \frac{1}{n-m+1} \langle\sum_{l=m}^{n} C_{l, m} \rangle \qc n=\mbox{min}(N_\varphi-1,(N_\theta+1)/2)
\end{equation}
(the angular parenthesis indicate time and radial average).

In the manuscript we use the terminology \textit{main-degree} when referring to $l_m=\argmax_{l\neq 0} (C_l)$, and \textit{main-order} for $m_m=\argmax_{m\neq 0} (C_m)$.

\bibliographystyle{jfm} 
\bibliography{biblio}

\begin{thebibliography}{32}
\expandafter\ifx\csname natexlab\endcsname\relax\def\natexlab#1{#1}\fi
\def\au#1{#1} \def\ed#1{#1} \def\yr#1{#1}\def\at#1{#1}\def\jt#1{\textit{#1}}
  \def\bt#1{#1}\def\bvol#1{\textbf{#1}} \def\vol#1{#1} \def\pg#1{#1}
  \def\publ#1{#1}\def\arxiv#1{#1}\def\org#1{#1}\def\st#1{\textit{#1}}

\bibitem[Adams \& Swarztrauber(1999)]{adams1999spherepack}
{\sc \au{Adams, J.~C.} \& \au{Swarztrauber, P.~N.}} \yr{1999}  \at{Spherepack
  3.0: A model development facility}.  \jt{Mon. Weather Rev.}  \bvol{127}~(8),
  \pg{1872--1878}.

\bibitem[Ahlers {\em et~al.\/}(2009)Ahlers, Grossmann \& Lohse]{Ahlers2009}
{\sc \au{Ahlers, G.}, \au{Grossmann, S.} \& \au{Lohse, D.}} \yr{2009}  \at{Heat
  transfer and large scale dynamics in turbulent {{Rayleigh-B{\'e}nard}}
  convection}.  \jt{Rev. Mod. Phys.}  \bvol{81},  \pg{503--537}.

\bibitem[Araki {\em et~al.\/}(1994)Araki, Mizushima \&
  Yanase]{araki1994thermal}
{\sc \au{Araki, K.}, \au{Mizushima, J.} \& \au{Yanase, S.}} \yr{1994}
  \at{Thermal instability of a fluid in a spherical shell with thin layer
  approximation analysis}.  \jt{Journal of the Physical Society of Japan}
  \bvol{63}~(6),  \pg{2123--2132}.

\bibitem[Avila {\em et~al.\/}(2013)Avila, Cabello-Gonz{\'a}lez \&
  Ramos]{Avila2013}
{\sc \au{Avila, R.}, \au{Cabello-Gonz{\'a}lez, A.} \& \au{Ramos, E.}} \yr{2013}
   \at{A linear stability analysis of thermal convection in spherical shells
  with variable radial gravity based on the tau-chebyshev method}.
  \jt{International journal of heat and fluid flow}  \bvol{44},  \pg{495--508}.

\bibitem[Bercovici {\em et~al.\/}(1989)Bercovici, Schubert \&
  Glatzmaier]{bercovici1989three}
{\sc \au{Bercovici, D.}, \au{Schubert, G.} \& \au{Glatzmaier, G.~A.}} \yr{1989}
   \at{Three-dimensional spherical models of convection in the earth's mantle}.
   \jt{Science}  \bvol{244}~(4907),  \pg{950--955}.

\bibitem[Bishop {\em et~al.\/}(1966)Bishop, Mack \& Scanlan]{bishop1966heat}
{\sc \au{Bishop, E.}, \au{Mack, L.} \& \au{Scanlan, J.}} \yr{1966}  \at{Heat
  transfer by natural convection between concentric spheres}.  \jt{Int. J. Heat
  Mass Transfer}  \bvol{9}~(7),  \pg{649--662}.

\bibitem[Busse(1970)]{busse1970differential}
{\sc \au{Busse, F.}} \yr{1970}  \at{Differential rotation in stellar convection
  zones}.  \jt{Astrophys. J.}  \bvol{159},  \pg{629}.

\bibitem[Busse(1975)]{busse1975patterns}
{\sc \au{Busse, F.}} \yr{1975}  \at{Patterns of convection in spherical
  shells}.  \jt{J. Fluid Mech.}  \bvol{72}~(1),  \pg{67--85}.

\bibitem[Chandrasekhar(1961)]{chandrasekhar1961hydrodynamic}
{\sc \au{Chandrasekhar, S.}} \yr{1961}  \at{Hydrodynamic and hydromagnetic
  stability}.  \jt{International Series of Monographs on Physics} .

\bibitem[Chilla \& Schumacher(2012)]{Chilla2012}
{\sc \au{Chilla, F.} \& \au{Schumacher, J.}} \yr{2012}  \at{New perspectives in
  turbulent {{Rayleigh-B{\'e}nard}} convection}.  \jt{Eur. Phys. J. E}
  \bvol{35},  \pg{58}.

\bibitem[Feldman {\em et~al.\/}(2012)Feldman, Colonius, Pauken, Hall \&
  Jones]{feldman2012simulation}
{\sc \au{Feldman, Y.}, \au{Colonius, T.}, \au{Pauken, M.~T.}, \au{Hall, J.~L.}
  \& \au{Jones, J.~A.}} \yr{2012}  \at{Simulation and cryogenic experiments of
  natural convection for the titan montgolfiere}.  \jt{AIAA journal}
  \bvol{50}~(11),  \pg{2483--2491}.

\bibitem[Futterer {\em et~al.\/}(2008)Futterer, Gellert, Von~Larcher \&
  Egbers]{futterer2008thermal}
{\sc \au{Futterer, B.}, \au{Gellert, M.}, \au{Von~Larcher, T.} \& \au{Egbers,
  C.}} \yr{2008}  \at{Thermal convection in rotating spherical shells: An
  experimental and numerical approach within geoflow}.  \jt{Acta Astronautica}
  \bvol{62}~(4-5),  \pg{300--307}.

\bibitem[Futterer {\em et~al.\/}(2013)Futterer, Krebs, Plesa, Zaussinger,
  Hollerbach, Breuer \& Egbers]{Futterer2013}
{\sc \au{Futterer, B.}, \au{Krebs, A.}, \au{Plesa, A.-C.}, \au{Zaussinger, F.},
  \au{Hollerbach, R.}, \au{Breuer, D.} \& \au{Egbers, C.}} \yr{2013}
  \at{Sheet-like and plume-like thermal flow in a spherical convection
  experiment performed under microgravity}.  \jt{J. Fluid Mech.}  \bvol{735},
  \pg{647–683}.

\bibitem[Garg(1992)]{garg1992natural}
{\sc \au{Garg, V.~K.}} \yr{1992}  \at{Natural convection between concentric
  spheres}.  \jt{Int. J. Heat Mass Transfer}  \bvol{35}~(8),  \pg{1935--1945}.

\bibitem[Gastine {\em et~al.\/}(2015)Gastine, Wicht \& Aurnou]{Gastine2015}
{\sc \au{Gastine, T.}, \au{Wicht, J.} \& \au{Aurnou, J.~M.}} \yr{2015}
  \at{{Turbulent Rayleigh--B{\'e}nard convection in spherical shells}}.  \jt{J.
  Fluid Mech.}  \bvol{778},  \pg{{721–764}}.

\bibitem[Hart {\em et~al.\/}(1986)Hart, Glatzmaier \& Toomre]{hart1986space}
{\sc \au{Hart, J.~E.}, \au{Glatzmaier, G.~A.} \& \au{Toomre, J.}} \yr{1986}
  \at{Space-laboratory and numerical simulations of thermal convection in a
  rotating hemispherical shell with radial gravity}.  \jt{Journal of Fluid
  Mechanics}  \bvol{173},  \pg{519--544}.

\bibitem[Iwase \& Honda(1997)]{iwase1997interpretation}
{\sc \au{Iwase, Y.} \& \au{Honda, S.}} \yr{1997}  \at{An interpretation of the
  {{Nusselt-Rayleigh}} number relationship for convection in a spherical
  shell}.  \jt{Geophys. J. Int.}  \bvol{130}~(3),  \pg{801--804}.

\bibitem[Joseph \& Carmi(1966)]{joseph1966subcritical}
{\sc \au{Joseph, D.~D.} \& \au{Carmi, S.}} \yr{1966}  \at{Subcritical
  convective instability part 2. spherical shells}.  \jt{Journal of Fluid
  Mechanics}  \bvol{26}~(4),  \pg{769--777}.

\bibitem[Mack \& Hardee(1968)]{mack1968natural}
{\sc \au{Mack, L.~R.} \& \au{Hardee, H.~C.}} \yr{1968}  \at{Natural convection
  between concentric spheres at low {Rayleigh} numbers}.  \jt{Int. J. Heat Mass
  Transfer}  \bvol{11}~(3),  \pg{387--396}.

\bibitem[Mannix \& Mestel(2019)]{mannix2019weakly}
{\sc \au{Mannix, P.} \& \au{Mestel, A.}} \yr{2019}  \at{Weakly nonlinear mode
  interactions in spherical rayleigh--b{\'e}nard convection}.  \jt{Journal of
  Fluid Mechanics}  \bvol{874},  \pg{359--390}.

\bibitem[O'Farrell {\em et~al.\/}(2013)O'Farrell, Lowman \&
  Bunge]{o2013comparison}
{\sc \au{O'Farrell, K.~A.}, \au{Lowman, J.~P.} \& \au{Bunge, H.-P.}} \yr{2013}
  \at{{Comparison of spherical-shell and plane-layer mantle convection thermal
  structure in viscously stratified models with mixed-mode heating:
  implications for the incorporation of temperature-dependent parameters}}.
  \jt{Geophys. J. Int.}  \bvol{192}~(2),  \pg{456--472}.

\bibitem[Santelli {\em et~al.\/}(2020)Santelli, Orlandi \&
  Verzicco]{santelli2020finitedifference}
{\sc \au{Santelli, L.}, \au{Orlandi, P.} \& \au{Verzicco, R.}} \yr{2020}
  \at{{A finite--difference scheme for three--dimensional incompressible flows
  in spherical coordinates}}.  \jt{J. Comput. Phys.}  \pg{p. 109848}.

\bibitem[Sattinger(1979)]{sattinger1979group}
{\sc \au{Sattinger, D.}} \yr{1979}  \at{Group theoretic methods in bifurcation
  theory}.  \jt{Lecture Note in Mathematics} .

\bibitem[Siggia(1994)]{siggia1994high}
{\sc \au{Siggia, E.~D.}} \yr{1994}  \at{High {{Rayleigh}} number convection}.
  \jt{Annu. Rev. Fluid Mech.}  \bvol{26}~(1),  \pg{137--168}.

\bibitem[Spiegel(1971)]{spiegel1971convection}
{\sc \au{Spiegel, E.~A.}} \yr{1971}  \at{{Convection in stars I. Basic
  Boussinesq convection}}.  \jt{Annu. Rev. Astron. Astrophys}  \bvol{9}~(1),
  \pg{323--352}.

\bibitem[Svensmark(2007)]{svensmark2007cosmoclimatology}
{\sc \au{Svensmark, H.}} \yr{2007}  \at{Cosmoclimatology: a new theory
  emerges}.  \jt{Astronomy \& Geophysics}  \bvol{48}~(1),  \pg{1--18}.

\bibitem[Svensmark \& Friis-Christensen(1997)]{svensmark1997variation}
{\sc \au{Svensmark, H.} \& \au{Friis-Christensen, E.}} \yr{1997}  \at{Variation
  of cosmic ray flux and global cloud coverage—a missing link in
  solar-climate relationships}.  \jt{Journal of atmospheric and
  solar-terrestrial physics}  \bvol{59}~(11),  \pg{1225--1232}.

\bibitem[Tilgner(1996)]{tilgner1996high}
{\sc \au{Tilgner, A.}} \yr{1996}  \at{{High-Rayleigh-number} convection in
  spherical shells}.  \jt{Phys. Rev. E}  \bvol{53}~(5),  \pg{4847}.

\bibitem[Travnikov {\em et~al.\/}(2003)Travnikov, Egbers \&
  Hollerbach]{travnikov2003geoflow}
{\sc \au{Travnikov, V.}, \au{Egbers, C.} \& \au{Hollerbach, R.}} \yr{2003}
  \at{The geoflow-experiment on iss (part ii): Numerical simulation}.
  \jt{Advances in Space Research}  \bvol{32}~(2),  \pg{181--189}.

\bibitem[Verzicco \& Orlandi(1996)]{Verzi96}
{\sc \au{Verzicco, R.} \& \au{Orlandi, P.}} \yr{1996}  \at{A finite--difference
  scheme for three--dimensional incompressible flows in cylindrical
  coordinates}.  \jt{J. of Comp. Phys}  \bvol{123},  \pg{402--414}.

\bibitem[Yanagisawa \& Yamagishi(2005)]{yanagisawa2005rayleigh}
{\sc \au{Yanagisawa, T.} \& \au{Yamagishi, Y.}} \yr{2005}
  \at{{Rayleigh--B{\'e}nard convection in spherical shell with infinite Prandtl
  number at high Rayleigh number}}.  \jt{J. Earth Simulator}  \bvol{4},
  \pg{11--17}.

\bibitem[Zebib {\em et~al.\/}(1980)Zebib, Schubert \&
  Straus]{zebib1980infinite}
{\sc \au{Zebib, A.}, \au{Schubert, G.} \& \au{Straus, J.~M.}} \yr{1980}
  \at{Infinite prandtl number thermal convection in a spherical shell}.
  \jt{Journal of Fluid Mechanics}  \bvol{97}~(2),  \pg{257--277}.

\end{thebibliography}

\end{document}